\documentclass[12pt]{iopart}

\usepackage[utf8]{inputenc}
\usepackage{iopams}  
\usepackage{graphicx}  
\expandafter\let\csname equation*\endcsname\relax
\expandafter\let\csname endequation*\endcsname\relax
\usepackage{amsmath}

\usepackage[labelfont={small},subrefformat=parens,caption=false]{subfig}
\usepackage{amssymb,amsbsy,amsfonts,bm}
\usepackage{gensymb}
\usepackage{bbm}
\usepackage{url} 
\usepackage{nicefrac}
\usepackage{hyperref}
\usepackage{xcolor}

\usepackage{dcolumn}
\newcolumntype{d}[1]{D{.}{.}{#1}}

\usepackage{xparse}

\usepackage{booktabs}  
\usepackage{multirow}
\usepackage{makecell}
\usepackage{cellspace}
\usepackage{dsfont}

\graphicspath{{figures/}}
\usepackage{bayesmacros}

\begin{document}

\title[Bayesian chiral EFT]{Exploring Bayesian parameter estimation \\
for chiral effective field theory using nucleon-nucleon phase shifts}

\author{S. Wesolowski$^{1,2}$, R.~J.\ Furnstahl$^{2}$, J.~A. Melendez$^{2}$, D.~R.\ Phillips$^{3,4,5}$}

\address{$^{1}$ Department of Mathematics and Computer Science, Salisbury University,
         Salisbury, MD 21801, USA\\
        $^{2}$ Department of Physics, The Ohio State University, 
         Columbus, OH 43210, USA \\
         $^{3}$ Institute of Nuclear and Particle Physics and Department of Physics and
         Astronomy, Ohio University, Athens, OH 45701, USA\\
         $^{4}$ Institut f\"ur Kernphysik, Technische Universit\"at Darmstadt, 64289 Darmstadt, Germany\\
	$^{5}$ ExtreMe Matter Institute EMMI, GSI Helmholtzzentrum f{\"u}r Schwerionenforschung GmbH, 64291 Darmstadt, Germany
         }

\ead{scwesolowski@salisbury.edu, furnstahl.1@osu.edu, melendez.27@osu.edu, phillid1@ohio.edu}

\begin{abstract}
We recently developed a Bayesian framework for parameter estimation in general effective field theories.
Here we present selected results from using that framework to estimate parameters with a nucleon-nucleon (\NN) potential
derived using chiral effective field theory (\eft): the semi-local \NN\ potential of Epelbaum, Krebs, 
and Mei\ss ner (EKM). There are many \NN\ scattering data, up to high energies, 
and with rather small errors, so imposing a penalty for unnatural low-energy constants (LECs) usually has a small
effect on the fits. In contrast, we have found that including an estimate of higher orders
in \eft\ plays an important role in robust parameter estimation.
We present two case studies where our Bayesian machinery illuminates physics issues. 
The first involves the EKM potential at fourth order in the \eft\ expansion: the two-dimensional
posterior probability density function (pdf) for the fourth-order $s$-wave 
LECs obtained from the Nijmegen PWA93 phase shifts
indicates these parameters in the \NN\ potential are degenerate. 
We trace this feature of the pdf to the presence of an operator in the fourth-order \NN\
potential that vanishes on-shell. The second case study examines the stability of  LEC extractions as 
more data at higher energies are included in the fit. 
We show that as long as \eft\ truncation errors are
properly accounted for in the parameter estimation, the LEC values 
extracted using our Bayesian approach are not sensitive to the maximum
energy chosen for the fit.
Uncorrelated and fully correlated models for the truncation errors are compared, pointing the
way to the use of Gaussian processes to more generally model the correlation structure.
\end{abstract}

\section{Introduction} \label{sec:introduction}

Effective field theories (EFTs) summarize high-energy~(short-distance) physics
using a series of operators that respect the symmetries of the underlying
theory. The separation of scales between high energies~(short distances), where
details of that underlying theory become relevant, and low energies~(long distances), where processes
of interest take place, allows the formation of an expansion parameter 
$\Q \equiv \mu/\breakdown$, where $\mu$ is a low-energy scale in the theory and 
$\breakdown$ is the breakdown scale. If the EFT is working as expected 
a prediction at order~$\kord$
has a truncation uncertainty dominated by terms of order $\Q^{\kord+1}$, and so its predictions can be systematically improved.  

Chiral effective field theory (\eft) is the extension of chiral perturbation theory to few-nucleon systems. As such it is a double expansion in 
$p/\breakdown$ and $m_\pi/\breakdown$, where $p$ is the typical momentum of the process, and $m_\pi$ is
the pion mass. If \eft\ were set up as a canonical EFT the expansion would be for observables, but instead 
it has mainly been implemented for the nucleon-nucleon (\NN) and three-nucleon (\NNN) potential (see \cite{Bedaque:2002mn,Epelbaum:2008ga,Epelbaum:2012vx} for
reviews). That potential is then iterated using the Schr\"odinger---or, equivalently the Lippman-Schwinger---equation. In such a calculation
the \eft\ potential must be regulated. Results of a calculation with a fixed-order potential are known to be regulator dependent based on both
 formal arguments~\cite{Nogga:2005hy} and results
for binding in light nuclei and nuclear matter~\cite{Dyhdalo:2016ygz,Huth:2017wzw}. But, at least for some specific regulators, the \eft\ expansion for the potential
 appears to yield a convergent \eft\ expansion for \NN\ observables as well~\cite{Furnstahl:2015rha,Melendez:2017phj,Epelbaum:2014efa}. And recent results for few-body 
 observables with these potentials, although not all computed as yet with the three-body forces needed for consistency at N2LO and beyond, are also promising in this regard~\cite{Binder:2015mbz,Binder:2018pgl,Lonardoni:2017hgs,Lonardoni:2018nob,Epelbaum:2018ogq}. 
 
At a fixed order, any EFT has a finite number of free parameters, 
the so-called low-energy constants (LECs).
These will be natural-sized (i.e.\ of order unity) if physical scales have
been properly identified. LECs can sometimes be found
by matching the EFT to the underlying high-energy theory, but in nuclear physics they are
more commonly obtained by fitting the EFT to data.
In fact, while we will use the verb ``fit'' to describe this process throughout this paper, we 
advocate sampling the full posterior probability density function (pdf)
of the LECs when feasible, rather than just finding the most probable values
through optimization.
This emphasizes that ``fitting"  the LECs induces uncertainties in and correlations between them.
Propagation of these is part of the uncertainty quantification for any EFT prediction.

Traditional fitting procedures for \eft\ LECs use least-squares minimization, often
augmented by constraints such as naturalness~(see, e.g.~\cite{Epelbaum:2014efa}). 
For \eft,
predictions are then usually made by using the ``best-fit'' values for the LECs.
Uncertainties from fitting, when provided, 
are estimated using covariance approximations for the least-squares likelihood~(see~\cite{Carlsson:2015vda,Tellinghuisen2001} and \eqref{eq:err-prop-formula}). 
Propagation of those uncertainties then approximately 
incorporates in predictions both the uncertainty of, and the correlations
between, LECs.

In addition to errors from the LECs themselves, systematic uncertainty must
then be estimated: this includes the EFT's truncation error for the observable and uncertainties from numerical methods. 
In the past, EFT truncation errors have been obtained by taking the largest 
difference between predictions using a range of regulator cutoffs.
 However, error bands from cutoff variation
do not have a clear statistical interpretation. 
Truncation error estimates based on using lower-order EFT calculations to inform
the size of omitted terms have recently become more 
common~\cite{Epelbaum:2014efa,Gandolfi:2016bth,Reinert:2017usi};
in particular,~\cite{Epelbaum:2014efa} codified
a standard EFT uncertainty estimation protocol and applied it to \NN\ predictions. 
In \cite{Furnstahl:2015rha}
we showed that this protocol can be interpreted as part of a more general
Bayesian uncertainty analysis for estimating EFT truncation errors. 
This statistical approach to EFT truncation uncertainties was further developed in \cite{Melendez:2017phj}.

Progress toward full uncertainty quantification continues to be made for modern \eft\ 
 potentials~\cite{Carlsson:2015vda,Epelbaum:2014efa,Reinert:2017usi}, but a consistent approach 
 to account for all uncertainties has not been reached.
We advocate the use of Bayesian methods
for this task~\cite{Schindler:2008fh,Furnstahl:2014xsa,Wesolowski:2015fqa}:
here, parameter estimation is the process
by which experimental data are used to generate a joint posterior pdf for the LECs,
see, e.g.~Figs.~\subref*{fig:phaseshift_post_1P1_N2LO_Emax_100MeV} and \subref*{fig:phaseshift_post_1S0_NLO_Emax_100MeV}.  
From the LECs' posterior pdf, one can 
define a `best fit' value as the mean, median, or mode of the distribution, and the uncertainty
as its (co)variance.
EFT expectations regarding naturalness and truncation
errors are encoded in the Bayesian analysis through prior pdfs. 
Posteriors then have a structure that includes the effect of truncation errors 
and encodes all pertinent information regarding correlations between LECs.
We stress that the need to specify priors means that
all theoretical assumptions are explicit in the
calculation of the posterior, making such an analysis reproducible and testable. 
The impact of uncertainties and correlations on predictions can be straightforwardly obtained by 
sampling the posterior pdf and generating predictions from those samples.

In the presence of large quantities of data, parameter estimates from 
Bayesian posterior pdfs may become equivalent to those from standard optimization methods.
Here we will consider the \NN\ sector, where data are
highly precise and cover a large energy range~\cite{PhysRevC.48.792,Perez:2013mwa}. 
If multiple LECs are fit to a small energy range, 
prior information such as naturalness will substantially 
influence LEC fits (see Table~\ref{tab:prior-compare-LECs-1P1-N3LO}), but this
influence goes away with a large energy range.
However, Bayesian methods yield other advantages, in particular:
(1) LEC posterior pdfs can be mined to uncover physics issues previously overlooked, 
and (2) modelling the contributions from omitted terms in the EFT expansion makes parameter estimates 
stable as the energy range of the fit is varied and yields consistent uncertainties. 
In this work we explore these advantages of LEC parameter estimation for 
\eft\ in the \NN\ sector using selected Bayesian graphical diagnostics. 
In future work we will show how the correct propagation of all \eft\ uncertainties to predictions of 
observables is facilitated within this framework.

Our explorations here are in the context of the (semi-local coordinate-space)
``EKM interactions" described in 
\cite{Epelbaum:2014efa,Epelbaum:2014sza}. The EKM \eft\ potential
	is specified in terms of a regularization procedure,
a well-defined power counting, and a documented fitting procedure.
EKM provide five order-by-order fits, distinguished by the choice
of regulator parameter, each up to fifth order in the chiral expansion.
Using Bayesian model-checking diagnostics, results in~\cite{Melendez:2017phj}
showed a systematic convergence pattern for observables for a particular
choice of regulator parameter ($R=0.9\,$fm, see \cite{Epelbaum:2014efa}). 
The consequent possibility of order-by-order comparisons
for this EKM interaction makes it an ideal test case for our framework.

Using our own code to compute \NN\ phase shifts from this potential, we fit the
LECs to the Nijmegen partial-wave analysis (PWA93)~\cite{PhysRevC.48.792}, which was used by
EKM in their fits. 
We summarize the data that will be used and the relevant formulas for calculating posterior pdfs 
in section~\ref{sec:setup}.  
Our method uses the convergence pattern
of observables and takes advantage of EFT properties 
to regulate the problem of LEC inference and prevent overfitting. 
The formalism is applicable not only to \eft, but can be used
for general EFT problems.

We then present two case studies for our framework. 
In section~\ref{sec:posteriors} we
 use projected posterior plots to display the information contained 
in the full posterior pdf for LECs. We use our results to elucidate some features of
the parameter estimation in the EKM interaction. Especially interesting is the result for $s$-wave parameters at fourth order in the \eft\ potential (\NNNLO), where 
the projected posterior pdf indicates a parameter degeneracy~\cite{Wesolowski:2016int}.
In fact, the two LECs implicated as degenerate by the posterior reduce to a single combination if the potential is evaluated in on-shell kinematics \footnote{This phenomenon was also studied in the recent chiral interaction of
Reinert, Krebs, and Epelbaum~\cite{Reinert:2017usi}.}. We argue that only one combination of these fourth-order \eft\ LECs affects the on-shell amplitude at this order---we show this explicitly in a model which includes only short-range pieces of the \eft\ potential in \ref{app:redundantpionless}. 

Because this first issue occurs at fourth order in \eft, truncation errors do not play a large role in
identifying that particular problem, but in general  
truncation errors play a key role in EFT parameter estimation. 
References~\cite{Schindler:2008fh,Wesolowski:2015fqa} sought to account for the impact of higher-order terms on the extracted LECs 
 by placing Bayesian priors on the omitted coefficients in the EFT expansion. 
 In particular, it was shown that such a treatment yields LECs
 that are constant within uncertainties as more data at higher energies are included in the fit. 
 In section~\ref{sec:Emaxplots} we apply these ideas to \eft. We look at LEC extractions in
the \onePone\ channel at \NLO\ and \NNLO, in the \threePone\  channel at \NNNLO, and in 
the \oneSzero\ channel at NLO, comparing uncorrelated and fully correlated models for the truncation errors. 
In all cases we find that a proper Bayesian treatment of truncation errors produces values for the LECs that are stable 
with respect to the maximum energy chosen, within 
uncertainties that account for this
theory error. 
The
comparison between the limiting models for the correlation structure 
motivates a more general Gaussian process model~\cite{GPpaper:2018}.

Section~\ref{sec:summary} presents a summary and outlook.
Three appendices include details that help to make this paper self-contained. 
In~\ref{sec:EFT-setup} we briefly review the general elements of EFTs
and specific details of \eft\ that are relevant for our study.
\ref{app:bayes} provides an overview of the Bayesian methods 
needed for parameter estimation, the priors we employ, and
the derivation of the posterior pdf formulas that are used for parameter estimation,
building on the developments and tests 
in \cite{Schindler:2008fh,Furnstahl:2014xsa,Wesolowski:2015fqa}. 
Finally, \ref{app:redundantpionless} provides additional arguments pertinent to the issue of short-range operator redundancy in the \NNNLO\ \eft\ potential.

\section{Formulas, software and data} \label{sec:setup}

In this section we first summarize the posterior
pdfs used for our calculation and specify our choice of priors in section~\ref{subsec:posterior_pdf}. 
Section~\ref{sec:ourcalc} describes some details of our calculations and section~\ref{sec:furtherdetails}
outlines our prescription for assigning uncertainties to phase shifts.

\subsection{Posterior pdf for LECs} \label{subsec:posterior_pdf}

The Bayesian parameter estimation of the $\kord$th order LECs $\LECk$ is based on the
posterior $\pr(\LECk \given \genobsvecexp, I)$, where $\genobsvecexp$ is a set of experimental 
measurements of an observable $y$ 
at $\Nexp$ kinematic points, and $I$ stands generically for information taken as given.
We will specify $I$ more completely as we proceed.
In this work $\genobsvecexp$ will be phase shifts or partial-wave cross sections at a set of energies in a fixed partial wave, but
the formalism applies to general data such as a mix of total or differential cross sections and spin observables.
An expression for the posterior is derived based on an underlying statistical model for $y$,
\begin{align}
    \label{eq:stat_model}
    \genobsvecexp = \genobsvecth + \Delta\genobsvecth + \Delta\genobsvecexp \;,  
\end{align}
where the $\kord$th order theoretical calculations $\genobsvecth$ at the $\Nexp$ points are a
function of the LECs at that order:
\begin{align}
  \genobsvecth \longrightarrow \genobsveck(\LECk)
  \;.
\end{align}
The experimental uncertainty is modelled by $\Delta\genobsvecexp$, which includes stochastic fluctuations
as well as systematic errors (e.g.\ normalization uncertainties).
The theory discrepancy term $\Delta\genobsvecth$ is often ignored during EFT fitting and prediction, but
here we introduce a model for this uncertainty due to higher-order EFT contributions---in the statistics
literature this is sometimes referred to as a ``model discrepancy"~\cite{KennedyBayesiancalibrationcomputer2001}. A more general model discrepancy 
function could also include the uncertainty from the calculational method.

In Bayesian statistics, unknown quantities, such as the experimental and
theoretical uncertainty, are treated as random variables. 
The path from the random variables in \eqref{eq:stat_model} to the posterior 
$\pr(\LECk \given \genobsvecexp, I)$ is traced out in \ref{app:bayes}.
In the present analysis we choose prior pdfs and make some simplifying assumptions such that all of the pdfs 
are normal distributions.
This enables clear and intuitive formulas while leaving open the possibility of other assumptions. 

Assuming a Gaussian form for both the theory and experimental uncertainties (see below
and Appendix B), it follows from \eqref{eq:stat_model} that the posterior for the LECs takes the general form
\begin{align}  \label{eq:full_posterior_for_LECs}
   \pr(\LECk \given \genobsvecexp, \covarexp, \covarth) & \propto
    \pr(\genobsvecexp \given \genobsvecth,  \covarexp, \covarth) \, \pr(\LECk \given \abar)
    \notag \\
   & \propto
   \eup^{-\frac{1}{2}\resvec^\trans \cdotbold (\covarexp + \covarth)^{-1} \cdotbold \resvec }
   \,
   \eup^{-(\LECk)^2/2\abar^2} 
  \;,
\end{align}
where the residual $\resvec$ is defined as
\begin{align}  \label{eq:residual_def}
   \resvec \equiv \genobsvecexp - \genobsvecth 
   \;.
\end{align}
Any quantities in~\eqref{eq:full_posterior_for_LECs} and~\eqref{eq:residual_def} that
depend on $\genobsvecth$ or $\covarth$ implicitly depend on the order $\kord$.
The parameters that will determine the posterior are summarized in Table~\ref{tab:parameters_in_posterior}.
The last factor in \eqref{eq:full_posterior_for_LECs} is a prior for $\LECk$ based on a naturalness assumption: if the relevant
physical scales are identified, the scaled (dimensionless) LECs should all be about the same magnitude.
We model the $a_i$s as statistically independent with a distribution characterized by a parameter $\abar$.
For clarity we have taken a fixed value
of $\abar$, which yields the Gaussian in \eqref{eq:full_posterior_for_LECs} (see \ref{app:bayes}),
and have chosen $\abar=5$ for the numerical results.
In general, however, we advocate marginalizing over $\abar$ with an appropriate prior as in \eqref{eq:abar_marginalization}.  
The manifest role
of this term is to impose a penalty on any LEC that is too large, which generally indicates overfitting.
For fitting phase shifts, this is only an issue if too many LECs are being fit to too small a range of
data, as illustrated in Table~\ref{tab:prior-compare-LECs-1P1-N3LO} in section~\ref{sec:post-information}.

We model $\Delta\genobsvecexp$ as independent Gaussian noise at each kinematic point $i$ with
zero mean and standard deviation $\sigmaexp_i$ (see \eqref{eq:prior_for_data}), which defines the 
covariance matrix $\covarexp$,
\begin{align} \label{eq:exp_covariance}
   (\covarexp)_{ij} = \sigmaexp_i^2 \delta_{ij}
   \;.
\end{align}
By our assumption the covariance matrix for the experimental error is diagonal,
but more generally it has non-zero off-diagonal elements if experimental data are correlated.
In that case, one can simply substitute an appropriate covariance matrix.
If the theory errors were negligible and we also neglected the naturalness penalty term, 
then \eqref{eq:full_posterior_for_LECs} would be
the conventional $\eup^{-\chi^2/2}$ likelihood given by the sum of the squared residuals weighted
by an experimental variance.

\begin{table}[t!]\centering
  \caption{Parameters determining the posterior \eqref{eq:full_posterior_for_LECs}.}
  \vspace{1mm}
  \renewcommand{\arraystretch}{1.15}
  \begin{tabular}{ll}
  \br
    $\kord$ & \quad chiral EFT expansion order (see \ref{sec:EFT-setup}); \\
    $\kmax$ & \quad highest order for omitted EFT contributions ($\kmax \ge k+1$);\\
    $\Nexp$ & \quad number of kinematic points used in a fit (size of boldface vectors);\\
    $\genobsvecexp$ & \quad Nijmegen PWA93 data in the pertinent partial wave;\\
    $\sigmaexpvec$ & \quad experimental error at each kinematic point; \\
    $\Qvec$ & \quad expansion parameter $Q_i = \{p_i,m_\pi\}/\Lambda_b$ (see \ref{sec:EFT-setup}); \\
    $\genobsrefvec$ & \quad reference scale for each observable at each kinematic point; \\
    $\LECk$ & \quad set of LECs at order $k$ in the chiral EFT expansion; \\
    $\Emax$ & \quad the largest lab energy for which we take phase shifts in $\genobsvecexp$;\\
    $\abar$ & \quad expected LEC size (this encodes naturalness of the LECs);\\
    $\cbar$ & \quad expected size of coefficients in the EFT series for observables 
        ($\crms$ here) \\
    \br
  \end{tabular}
  \label{tab:parameters_in_posterior}
\end{table}

We assume the theory error is dominated by
the EFT expansion truncation error, which has been modelled in 
\cite{Furnstahl:2015rha,Melendez:2017phj,GPpaper:2018}.
In particular, order-by-order calculations at kinematic point $i$, together with an expansion parameter $Q_i$ 
and a reference scale $(\genobsrefvec)_i$ (here assumed to be given, see \ref{sec:EFT-setup} and \ref{app:bayes}), are used to define expansion coefficients $c_{n}$
(we suppress the $i$ index on $c_n$ for clarity),
\begin{align} \label{eq:coefficient_expansion}
  [\genobsveck(\LECk)]_i = (\genobsrefvec)_i \sum_{n=0}^{k} c_{n} Q_i^n  
  \;. 
\end{align}
This means that the coefficients at order $n>0$ are found from calculations at two consecutive orders:
\begin{align}  \label{eq:coefficient_definition}
  c_{n} = c_{n}(\LECn,\LEC_{n-1}) 
        = \frac{[\genobsvecn(\LECn) - \genobsvec_{n-1}(\LEC_{n-1})]_i}{(\genobsrefvec)_i Q_i^n}
        \equiv
        \frac{[\Delta\genobsvecn]_i}{(\genobsrefvec)_i Q_i^n}
        \;.
\end{align}
The key expectation for a well-behaved EFT is that the $c_{n}$ defined this way are natural (of order unity).
The idea is that the correction at each successive order is roughly smaller by a factor $Q_i$
(and not that the calculation takes the functional form of \eqref{eq:coefficient_expansion}).
The truncation error is identified by the extension of \eqref{eq:coefficient_expansion} to 
order $\kmax$ 
(or summed to all orders),
\begin{align}  \label{eq:truncation_error}
  (\Delta\genobsvecth)_i 
  = (\genobsrefvec)_i \sum_{n=k+1}^{\kmax} c_{n} Q_i^n
  \;.
\end{align}
The model discrepancy function we adopt for chiral EFT is that the $c_n$ are independent and identically distributed (i.i.d.)
random variables, with a characteristic size $\cbar$.   

Experience with the coefficients $c_n$ for \NN\ scattering observables as a function of energy and/or angle
motivates two characterizations of their distributions~\cite{Furnstahl:2015rha,Melendez:2017phj}. 
In particular, the extent of the coefficient curves seems roughly independent of the kinematic point
in most cases, so we assume that the same $\cbar$ applies for all points. 
Its distribution is informed by the empirical
variance of the coefficients from \eqref{eq:coefficient_expansion}.
The values of the coefficients vary with kinematics in a fairly regular way, suggesting a
characteristic correlation length.  A model using Gaussian processes to capture this correlation
will be explored in \cite{GPpaper:2018}.
Here we consider the two extremes of very small and very large correlation length:
``uncorrelated'' means that coefficients at all kinematic points are treated as independent while ``fully correlated'' means
the coefficients at all points are the same.
The plots of $c_n$s extracted from observables in \cite{Melendez:2017phj} imply that neither 
 assumption is fully realistic for the meshes of energies we use in our fits---and
the distributions of coefficients are less regular in individual partial waves---but by considering
these two extremes of $c_n$s' behaviour with energy we can test the extent to which 
assumptions regarding that behaviour affect LEC parameter estimation.

Results for these two limits are derived for Gaussian priors in \ref{app:bayes}.
The covariance matrix for the theory error at order $k$ in the uncorrelated limit is diagonal,
\begin{align} \label{eq:covariance_case_A}
  (\covartharg{uncorr.})_{ij} =  (\genobsrefvec)_i^{2}\, \cbar^2 \sum_{n=k+1}^{\kmax} Q_i^{2n} \, \delta_{ij}
    \underset{\kmax\rightarrow\infty}{\longrightarrow} 
    \frac{(\genobsrefvec)_i^{2}\, \cbar^2 \, Q_i^{2k+2}}{1-Q_i^2} \, \delta_{ij}
  \;,
\end{align}
while in the fully correlated limit there are off-diagonal entries,
\begin{align} \label{eq:covariance_case_B}
  (\covartharg{corr.})_{ij} = 
   (\genobsrefvec)_i (\genobsrefvec)_j \, \cbar^2 \sum_{n=k+1}^{\kmax} Q_i^{n} Q_j^{n}
    \underset{\kmax\rightarrow\infty}{\longrightarrow} 
  \frac{(\genobsrefvec)_i (\genobsrefvec)_j \, \cbar^2 \, Q_i^{k+1} Q_j^{k+1}}{1-Q_i Q_j}
  \;.
\end{align}
If the expansion parameter is small then it is sufficient to take $\kmax = k+1$; we call this the ``first-omitted-term approximation".
In the present work we estimate $\cbar$ as 
$\sigmaemp$, the root-mean-square value of the expansion coefficients $c_n$ from calculations at a representative
sample of different energies (see \eqref{eq:crms_equation}).

The form of \eqref{eq:full_posterior_for_LECs} is maintained if the experimental errors become correlated or if other 
independent theory errors are included (in which case the latter becomes the sum of covariance matrices for
each error), \emph{if} they are all Gaussian. 
An earlier procedure by Carlsson et al.\ to account for truncation errors in parameter fits to \NN\ chiral EFT
used a likelihood function of the form \eqref{eq:mv_likelihood_quadrature}, which is \eqref{eq:full_posterior_for_LECs}
without the LEC prior, with a diagonal covariance matrix
for theory as in the uncorrelated limit~\cite{Carlsson:2015vda}:
\begin{align}
  (\covartharg{\mathrm{Ref}.{\mbox{\scriptsize\cite{Carlsson:2015vda}}}})_{ij} =  (C_x Q_i^{k+1})^{2} \, \delta_{ij}
  \;.
  \label{eq:covariance_Carlsson}
\end{align}
Thus this variance was added in quadrature as a penalty term, following the discussion in 
\cite{Dobaczewski:2014jga} of how to deal with systematic theory uncertainties.
The value of $C_x$ was determined by an iterative Birge factor procedure~\cite{Birge:1932}, 
in which $C_x$ was adjusted until the minimized $\chi^2$ per degree of freedom is close to unity
(recall that the likelihood is proportional to $\eup^{-\chi^2/2}$).

In previous work~\cite{Wesolowski:2015fqa}, we stated that ``The Bayesian approach we advocate 
for parameter estimation has a different structure to the procedures of 
\cite{Carlsson:2015vda,Epelbaum:2014efa}; it is an interesting and relevant question 
whether those procedures can also be derived or motivated by a Bayesian framework under
prescribed conditions.'' 
In that work and elsewhere~\cite{Schindler:2008fh,Furnstahl:2014xsa} the fully correlated
model for the truncation error corresponding to \eqref{eq:covariance_case_B} was incorporated
in the form of a modified, augmented $\chi^2$~\cite{Stump:2001gu} different from the
sum of variances in \eqref{eq:full_posterior_for_LECs}.
This alternative form is derived at the end of \ref{subsec:appendix_posterior_pdf}.
However, the form of the covariance matrix \eqref{eq:covariance_Carlsson} results from \eqref{eq:full_posterior_for_LECs}
and \eqref{eq:covariance_case_A},
so the Carlsson et al.\ procedure can be interpreted as incorporating a truncation error that is an independent Gaussian random variable at each data point (i.e.\ the uncorrelated limit)---provided we also use the first-omitted-term
approximation. 
Errors \emph{do} add in quadrature under these assumptions.
However, the Carlsson et al.\ procedure to determine $C_x$ imposes the self-consistent requirement that the mode of the
distribution of $\chi^2$ be close to what is expected from the number of degrees of freedom.
Using the $\chi^2$ per degree of freedom as an assessment of goodness-of-fit
for nonlinear parameter estimation is in general problematic (e.g.\ see \cite{Andrae:2010gh}).
Thus, we argue that the size of $C_x$, and hence of the 
truncation error, should be based solely on EFT naturalness.

\subsection{Our calculation} 
\label{sec:ourcalc}

The calculation of posterior pdfs requires a software pipeline for computing \NN\ observables
as the LECs are varied. We have developed a suite of codes that can use any EFT
interactions to calculate the predictions $\genobsvecth$ that appear in, 
e.g.~\eqref{eq:full_posterior_for_LECs}. The pipeline employs Markov
Chain Monte Carlo (MCMC) sampling, for which we use the \texttt{emcee} package~\cite{Foreman_Mackey:2013aa}
in \texttt{Python}. \NN\ observable calculations and posterior pdf evaluations 
are implemented in \texttt{C++}, but are called from \texttt{Python} as the MCMC sampling is performed.
This decouples the details of the pdf sampling from the evaluation of
observables, making it simple to include different processes (e.g.\ \piN\  scattering)
 in the data $\genobsvecexp$ in the future. 
Optimization and parallelization can be applied for the observable
codes and the MCMC sampling.
In what follows we show results with the $R=0.9$ fm version of the semi-local EKM potential of \cite{Epelbaum:2014efa} with $\Lambda_b = 600\,$MeV. 
Implementing other EFT interactions is straightforward, but we defer discussion of those to future work.

The end result of MCMC sampling is a representative set of samples, 
which can be histogrammed. It is straightforward to find the central value and
uncertainty of the LECs
from the samples, and they can be used to approximately evaluate
any integral over the LECs weighted by the LEC posterior. In this work,
when LEC credible intervals are quoted, we quote the median as the central
value and the 68\% interval between the 16th and 84th percentile as the uncertainty band.
However, if the LEC pdf is well-approximated by a Gaussian, simple covariance methods 
can instead reliably propagate
uncertainty from LECs to predictions. 
In what follows we use covariance methods by default.
The general formula for propagation of uncertainties
from LECs to some observable $y$ using a covariance approximation is
\begin{align}
	\sigma_y^2 = \mathbf{g}^T \mathbf{\Sigma} \mathbf{g} \;,
	\label{eq:err-prop-formula}
\end{align}
where $g_i = \partial y/\partial a_i$ and $\mathbf{\Sigma}$ is the 
covariance matrix of the parameters, all evaluated at the optimum of the objective function~\eqref{eq:full_posterior_for_LECs}. 
By encoding correlations and uncertainties
in the form of a covariance matrix, the coefficients are assumed to follow a multivariate
Gaussian pdf. 
We caution that the assumptions made in
\eqref{eq:err-prop-formula} may not always be sufficient
to capture the structure in the LEC pdf~\cite{Furnstahl:2014xsa}.

In order to check that our pipeline is robust, we generated synthetic phase shift data
 from the
EKM interactions by adopting the central values of LECs EKM extracted in their fits, computing phase shifts, and adding uncorrelated Gaussian noise. 
 We then took the
resulting output as input ``data'' for the pipeline. 
The resulting LEC posteriors were always centred at their input values.
Achieving this was a little more complicated at fourth order (\NNNLO) 
in the $s$-waves, since there the LEC
posterior is multi-modal even for synthetic data, and the sampling must be started
very close to the input EKM fit values to find the ``correct'' mode. We discuss this further 
in section~\ref{sec:swave-case}, where we explore an operator redundancy at
fourth order that can cause unexpected behaviour in the posterior pdf.

 \subsection{Specification of phase shifts, uncertainties, and energy range for parameter estimation}
 \label{sec:furtherdetails}

We estimate LECs of EKM's chiral interaction from the $np$ scattering phase shifts obtained 
in different partial waves in the partial wave analysis (PWA93) of the Nijmegen group [20].  
Larger data bases,  some with better uncertainty quantification,
exist~\cite{Perez:2013mwa}, but here we imitate EKM by using the PWA93.
For isoscalar channels the uncertainties
used in this section will be the statistical uncertainties reported for those channels in 
\cite{PhysRevC.48.792}.  These uncertainties, obtained by analysing
the statistical variation of boundary condition parameters in each partial wave
during the PWA93,
are not true experimental uncertainties. Because statistical uncertainties
are not provided in the PWA93 for isovector channels, for the ${}^3P_0$ channel in 
section~\ref{sec:post-information} we simply
assign the same uncertainties as used in the ${}^1P_1$ channel.
 Without true data uncertainties the
widths of the LEC posterior pdfs are determined by the size of the uncertainties we adopt.
A proper treatment of uncertainties
on phase shifts would treat them as correlated, model-dependent extractions from 
the \NN\ scattering data. However, we do not pursue that here and simply take statistical
uncertainties on phase shifts as independent.  While this is not adequate for 
a definitive extraction of the LECs in the \eft\ \NN\ potential, it is sufficient 
for the issues we are examining here: how correlations reveal underlying physics,
the stability of LEC extraction with $\Emax$, 
and how including truncation uncertainty affects the fits at different orders.
  
For fits to the phase shifts in section~\ref{sec:posteriors}, we take the same mesh of energies
used by EKM: $\Elab = 1,5,10,25,50,100,150,200\,\text{MeV}$. For reference,
the nominal values of $\Emax$ in the EKM fits~\cite{Epelbaum:2014efa} are:
LO, $\Emax=25\,\text{MeV}$; NLO and \NNLO, $\Emax=100\,\text{MeV}$; \NNNLO\ and \NNNNLO,
$\Emax=200\,\text{MeV}$. Restricting $\Emax$ is one way to account for the systematic
degradation of the EFT since higher-order terms of the EFT contribute more as the energy, and hence the expansion
parameter, increases. When doing parameter estimation in section \ref{sec:posteriors} we generally
use these values of $\Emax$, and state explicitly when different values were needed to reproduce EKM's LECs. However, 
we will argue
in section \ref{sec:Emaxplots} that the question of what $\Emax$ to choose is rendered moot by a proper treatment
of truncation errors. 

We also use slightly different data and a different prescription for the data errors in section~\ref{sec:Emaxplots}. 
In that section we incorporate truncation errors in our analysis; this necessitates an assessment of the convergence pattern of observables, 
and so we switch from fitting phase shifts to fitting (and predicting) partial-wave cross sections, denoted $\sigmapw$. The $\sigmapw$s 
can be easily computed from the 
phase shifts.  We do this on a finer mesh of lab energies: $\Elab = 1,5,10,25,50,75,100,125,150,175,200,225,250,275,300\,\text{MeV}$. Following EKM, we assign experimental errors to these cross sections
by taking the largest difference between different model potential predictions of $\sigmapw$ from 
the NN-online database. This leads to markedly larger uncertainties than are obtained by the prescription, described above, that is implemented
in section \ref{sec:posteriors}.

\section{Case study 1: the usefulness of projected posterior plots} \label{sec:posteriors}

The framework in~\cite{Wesolowski:2015fqa} consists of several distinct steps, starting
with the Setup, where all the input information, including the theory
itself, the prior assumptions, and the relevant data, are specified for the analysis,
then proceeding through Guidance, Parameter Estimation, Validation, and Predictions.
In this first case study, we focus on a key element of the Parameter Estimation stage, 
namely the projected posterior plot, which is a valuable tool for identifying and characterizing
multimodal behaviour and correlations between LECs.
We examine the posterior pdfs for LECs of the EKM interaction in various partial waves given the phase shifts
from the PWA93 database, first setting the stage in section~\ref{sec:post-information} 
with some characteristic examples of projected
posteriors to illustrate the information provided by these plots, and then considering 
in section~\ref{sec:swave-case} a
case where the projected posterior becomes a diagnostic for overfitting stemming from an
operator redundancy in the $s$-waves at \NNNLO. 
We postpone inclusion of truncation errors to section~\ref{sec:Emaxplots}, as they are less important
for this high-order example and do not affect the analysis of overfitting. 
Hence in this section we use $\Delta\genobsvecth = 0$, meaning the posterior pdf is
given by  \eqref{eq:full_posterior_for_LECs} with $\covarth$ set to zero.

\subsection{Information in projected posteriors} \label{sec:post-information}

\begin{figure*}[bht]
  \subfloat{%
    \label{fig:phaseshift_post_1P1_N2LO_Emax_100MeV}%
    \includegraphics[width=0.49\textwidth]{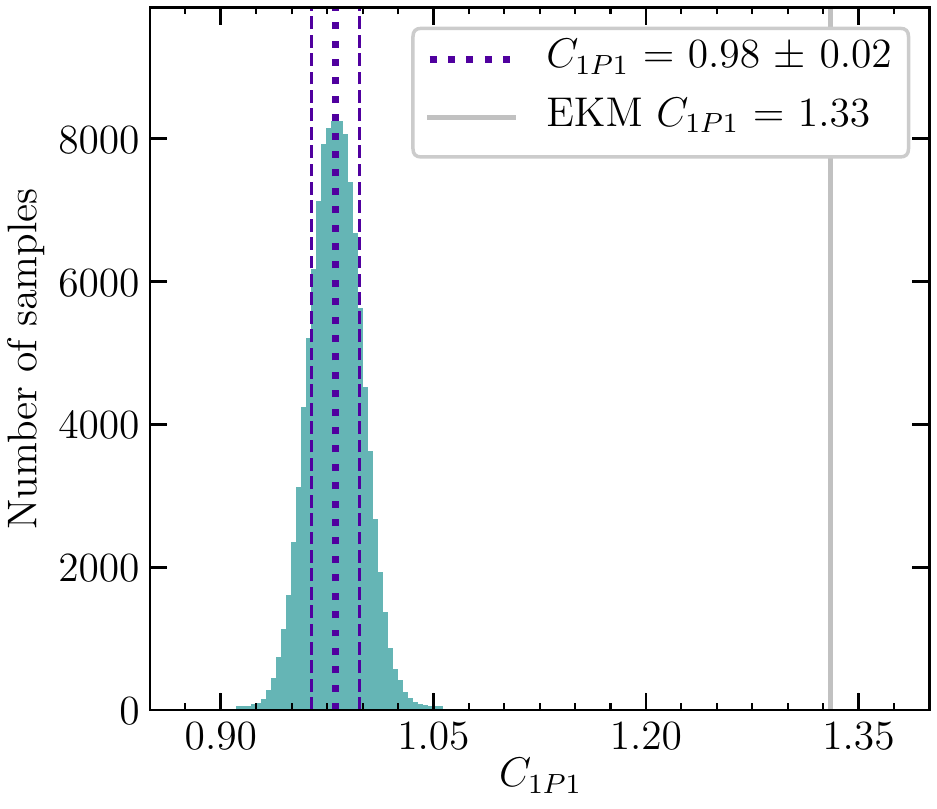}%
  }
  \hspace*{0.001\textwidth}  
  \subfloat{%
    \label{fig:phaseshift_1P1_N2LO_Emax_100MeV}%
    \includegraphics[width=0.49\textwidth]{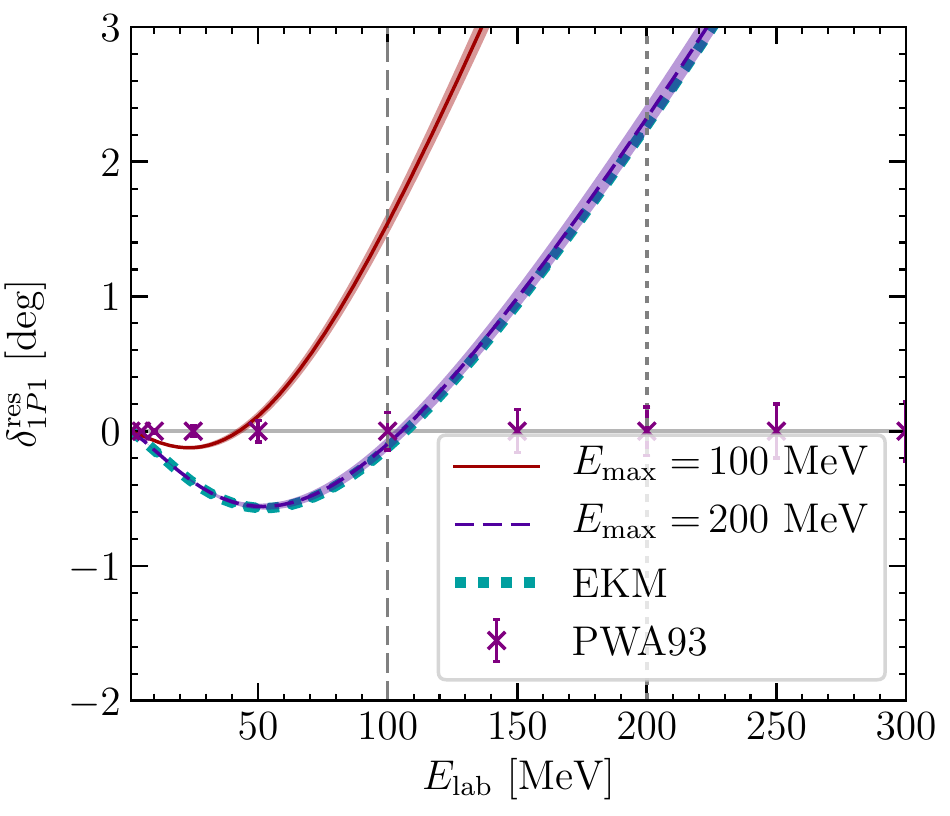}
  }
  \caption{(a) Histogram that approximates the (unnormalized) posterior pdf 
  for the \NNLO\ fit to PWA93 $np$ phase shifts 
  in the \onePone\  channel with $\Emax=100\mev$.
  The grey solid line shows the corresponding EKM value.
  (b) The red solid line and band show the 
  phase shift residuals \eqref{eq:pw_residual} compared to the PWA93 for the \onePone\ phase shift using the pdf 
  of $C_{1P1}$ with $\Emax = 100\mev$ from (a). 
  The purple dashed line and band are the residual and uncertainty obtained when $C_{1P1}$ is 
  estimated using data up to 
  $\Emax=200\mev$. (The dashed and dotted vertical lines indicate the two $\Emax$ limits). The teal dotted line is the EKM result for the residual.}
  \label{fig:both_phaseshift_1P1_N2LO_Emax_100MeV}
\end{figure*}

Figure~\subref*{fig:phaseshift_post_1P1_N2LO_Emax_100MeV} shows the sampled histogram that
approximates the (unnormalized)
posterior pdf for the single LEC in the \onePone\  channel at \NNLO\ ($\kord = 3$) with $\Emax=100\mev$.
Using standard normality tests~\cite{chambers1983gmd} such as the normal 
probability plot, we verify that
this one-dimensional posterior pdf is well-approximated as a Gaussian distribution.
In general, normality must be verified in each parameter
before applying related approximations, e.g.~\eqref{eq:err-prop-formula}.
Multidimensional tests of normality also exist, but have not been utilized in this work~\cite{HarrisonValidationBayesianposterior2015}.

\begin{figure*}[tb]
  \subfloat{
    \label{fig:phaseshift_post_3P0_N3LO_Emax_200MeV}%
    \includegraphics[width=0.49\textwidth]{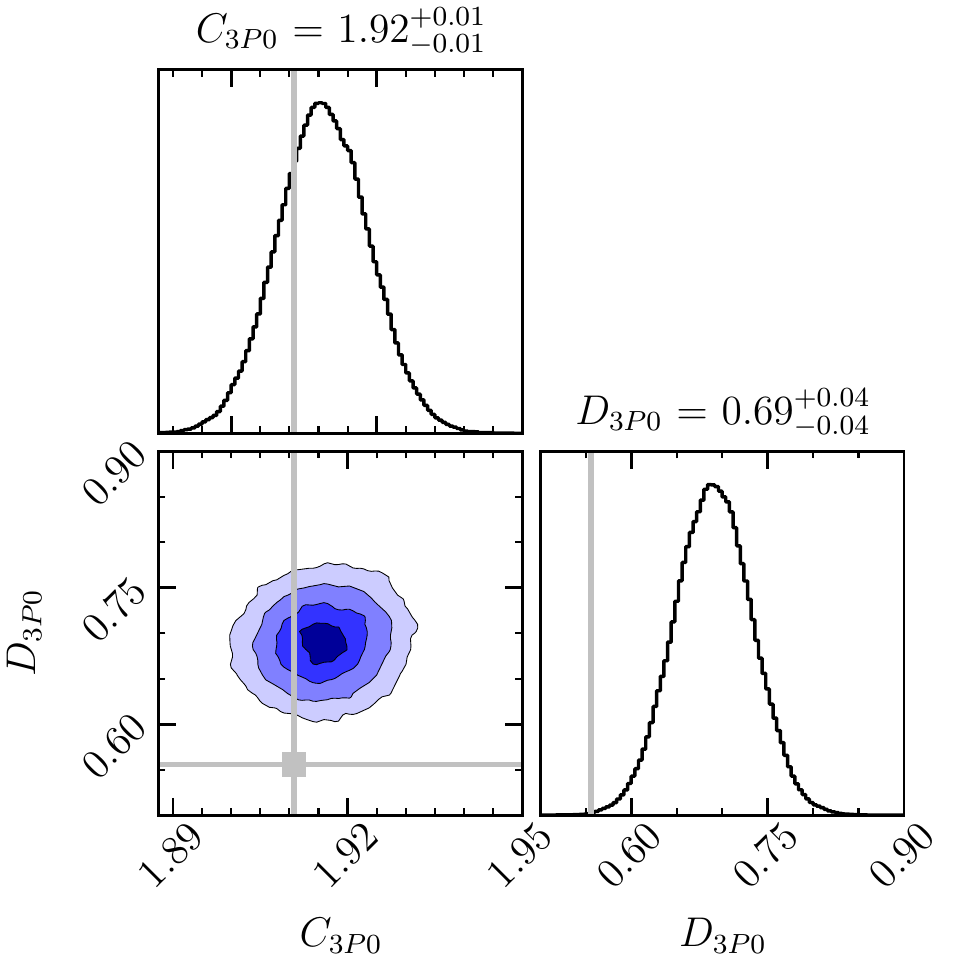}%
  }
  \hspace*{0.001\textwidth}  
  \subfloat{%
    \label{fig:phaseshift_3P0_N3LO_Emax_200MeV}%
    \includegraphics[width=0.49\textwidth]{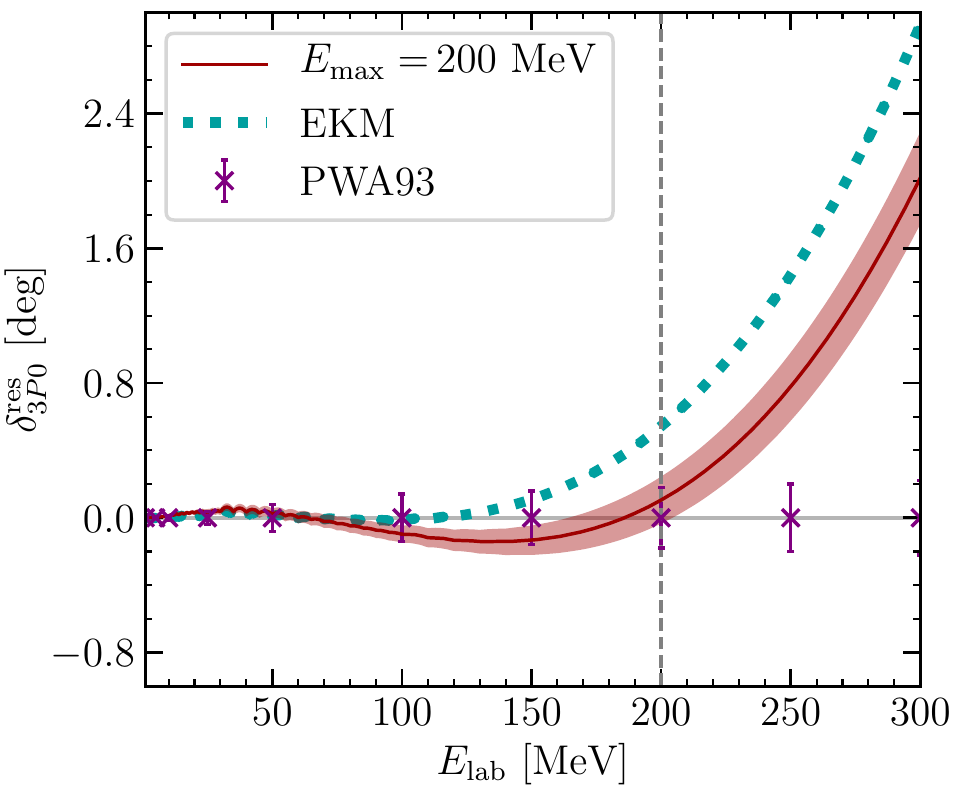}
  }
  \caption{(a) Full posterior in two dimensions (lower left) with one-dimensional projected posteriors (on the diagonal) for the 
  \NNNLO\ fit in the \threePzero\ channel 
  to PWA93 $np$ phase shifts with $E_{\mathrm{max}}=200\,$MeV.
  The contours show the $12\%$, $39\%$, $68\%$, and $86\%$ highest posterior density intervals, which correspond to $0.5\sigma$ increments up to $2\sigma$ when using a 2d Gaussian approximation.
  The grey solid lines and square show the corresponding EKM values. (b)
  The red solid line and band show the propagated phase shift residual
  using the joint pdf of LECs represented in (a), with $\Emax$
  indicated by the vertical dashed line. The teal dotted line is the EKM result. }
\end{figure*}

Whether the LEC's pdf is Gaussian or not, it is important to extract not only its 
most probable value but
also its uncertainty, so we can propagate LEC uncertainties to predictions for observables. 
(We reiterate
that the full uncertainty of an observable also includes the EFT truncation error 
and that the full uncertainty of an LEC should also include
the effect of higher-order terms on its estimation, cf.\ section \ref{sec:Emaxplots}.)
Figure~\subref*{fig:phaseshift_1P1_N2LO_Emax_100MeV} shows the 
phase-shift residual $\delta_{1P1}^{\text{res}}$, defined for a generic partial wave (p.w.) as
\beq
    \delta_{\text{p.w.}}^{\text{res}} = \delta_{\text{p.w.}} - \delta_{\text{p.w.}}^{\text{PWA93}}\;.
    \label{eq:pw_residual}
\eeq
The 68\% ($1\sigma$) band here and in the subsequent figures combines the uncertainty
in the LEC and the uncertainty in the phase shifts using~\eqref{eq:err-prop-formula}, although the effect of
the uncertainty in $C_{1P1}$ on the
\onePone\ phase-shift residual  
is very small for both $\Emax=100$ MeV and $\Emax=200$ MeV.

For $\Emax=100\,$MeV, the LEC extracted
from the PWA93 phase shifts is not
consistent with EKM's predictions at higher energies.
However, if we include data up to $\Emax=200\mev$, 
then we reproduce almost exactly the phase shifts of EKM, see figure~\subref*{fig:phaseshift_1P1_N2LO_Emax_100MeV}.
The difference between these predictions highlights the sensitivity to $\Emax$ and the need to address it,
which is the theme of our second case study in section~\ref{sec:Emaxplots}.
It is evident in figure~\subref*{fig:phaseshift_1P1_N2LO_Emax_100MeV} that neither fit exhibits
the steady degradation with energy expected for an EFT with correspondingly growing uncertainties;
instead we have underfitting and credible intervals inconsistent with the data errors.
We will need to add the truncation error in section~\ref{sec:Emaxplots} to avoid these
failings. 
Next we consider a case in which we have two \NN\ LECs, the
 \threePzero\  channel at \NNNLO\ ($\kord=4$) with $\Emax=200\,$MeV. 
In such cases, we look at a set of panels showing \emph{projected} posterior pdfs: the distribution is integrated
over various dimensions \cite{Wesolowski:2015fqa} 
to isolate the one-dimensional pdf of the LECs themselves in the diagonal panels, 
while the lower-left panel shows
the full two-dimensional posterior for $C_{3P0}$ and $D_{3P0}$, see figure~\subref*{fig:phaseshift_post_3P0_N3LO_Emax_200MeV}.
The projected posterior pdf yields not only the most likely values and uncertainties of individual LECs, but also the correlation between 
$C_{3P0}$ and $D_{3P0}$ that results from fitting them to the PWA93 data.
The untilted orientation of the two-dimensional credible-interval contours (called degree-of-belief or DoB
contours in previous work) show that these LECs are nearly uncorrelated.
Since we now have more than one
parameter we use a normal probability plot in each dimension to verify that the sampled
posterior is normal. 
Our extracted LECs and phase shift residuals at this nominal value of $\Emax=200\mev$ are close to EKM's results, but the value of $D_{3P0}$ disagrees with EKM
by significantly more than the LEC uncertainties.
The propagated phase shift residuals are shown and compared to the EKM results
in figure~\subref*{fig:phaseshift_3P0_N3LO_Emax_200MeV}. 

\begin{figure*}[ptbh]
  \subfloat{%
    \label{fig:phaseshift_post_1S0_NLO_Emax_100MeV}%
    \includegraphics[width=0.49\textwidth]{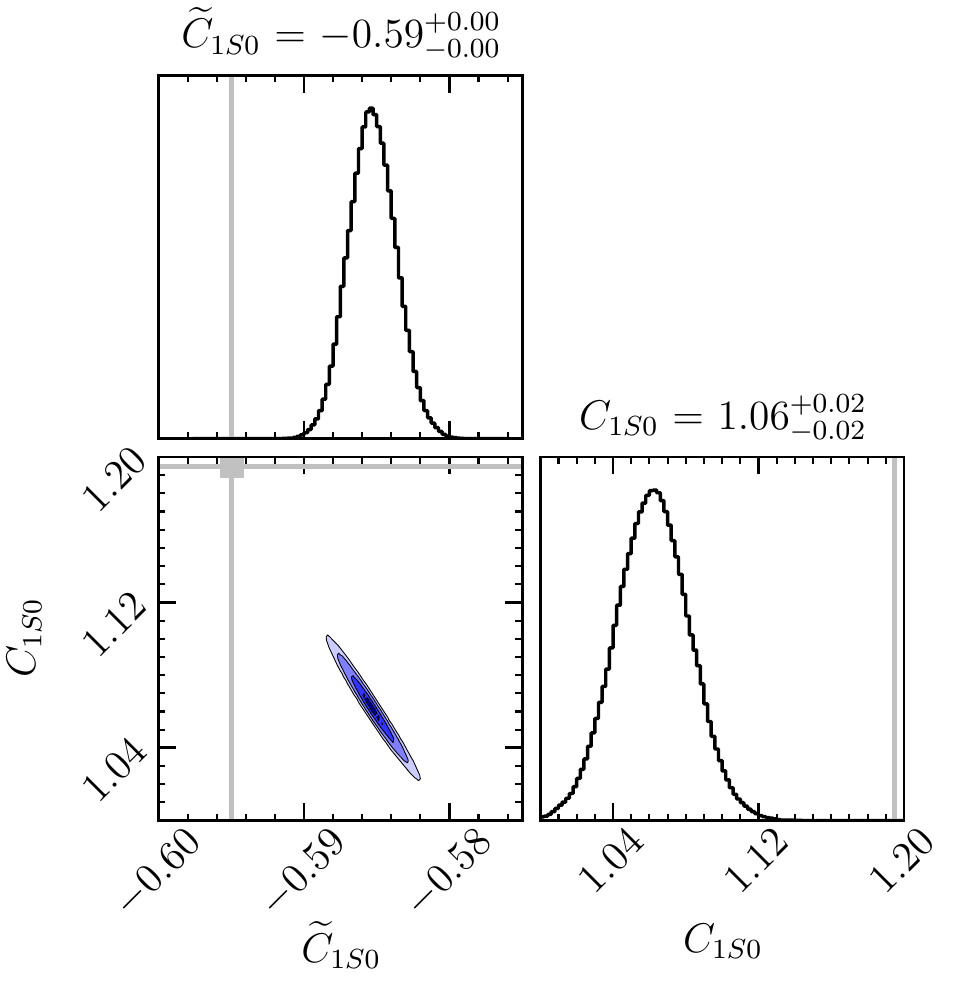}%
  }
  \hspace*{0.001\textwidth}  
  \subfloat{%
    \label{fig:phaseshift_1S0_NLO_Emax_100MeV}%
    \includegraphics[width=0.49\textwidth]{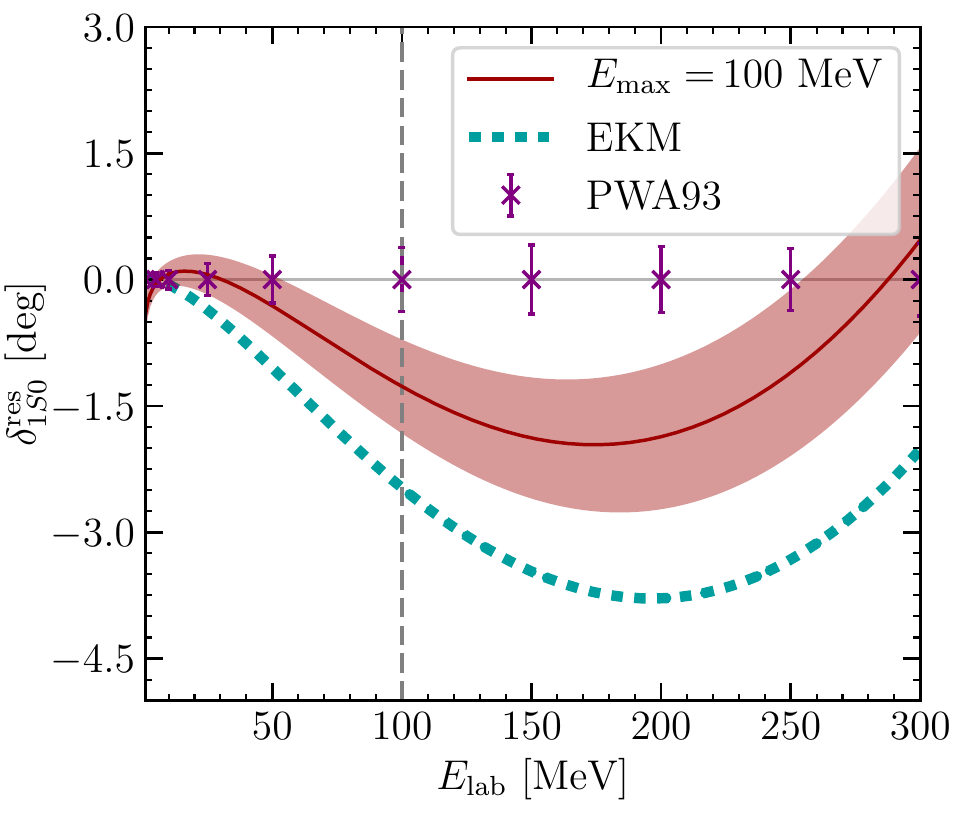}
  }
  \caption{(a) Posterior for the NLO fit in the \oneSzero\  channel 
  to the PWA93 $np$ phase shifts with $E_{\mathrm{max}}=100\,$MeV.
  The grey solid lines and square show the corresponding EKM values. (b)
  The red solid line and band show the propagated phase shift residual using the
  joint pdf of LECs represented in (a), with $\Emax$
  indicated by the vertical dashed line. The teal dotted line is the result with the EKM LECs.}
\end{figure*}
\begin{figure*}[pbth]
    \centering
	\includegraphics[width=0.60\textwidth]{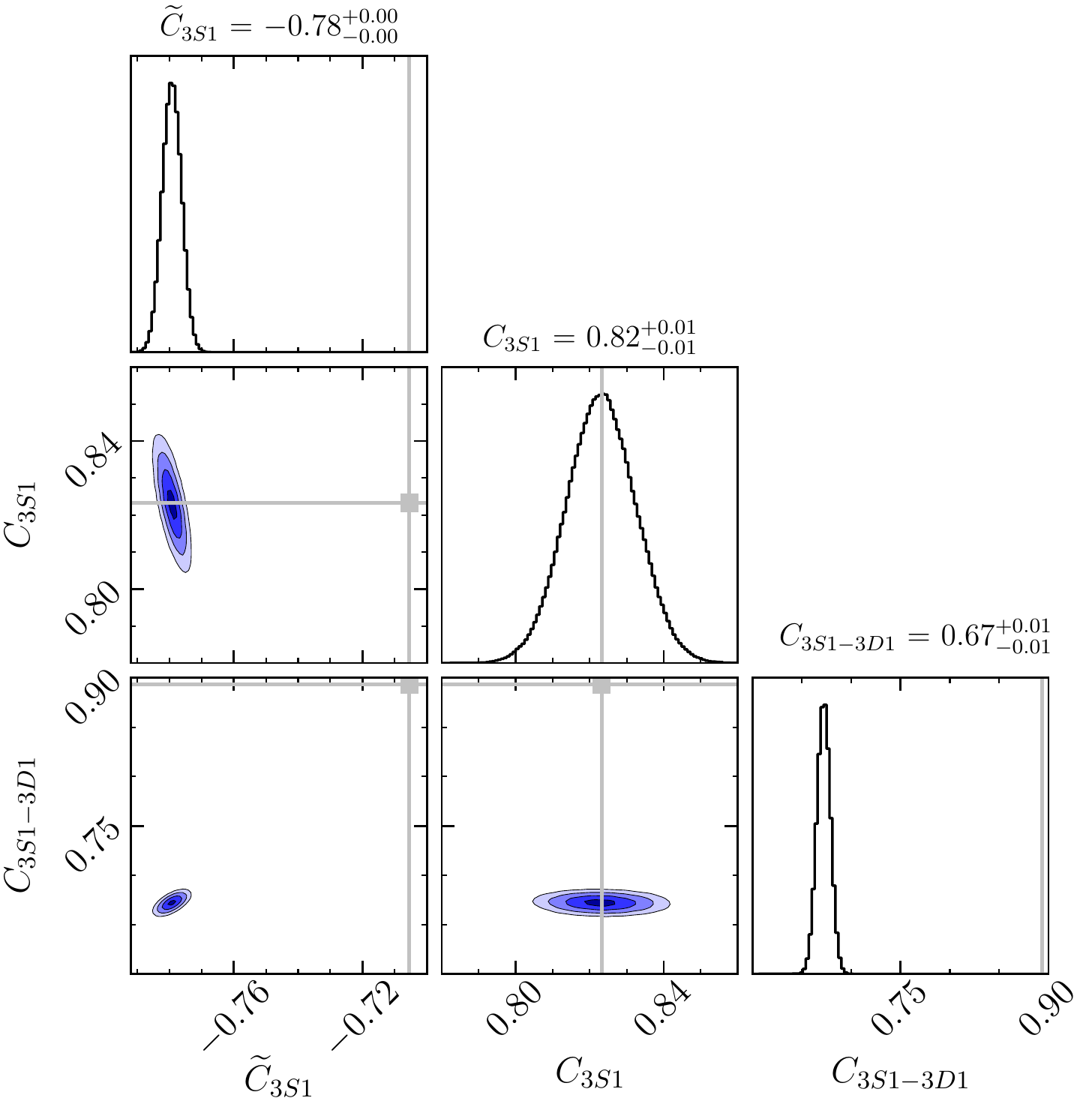}~%
    \includegraphics[width=0.37\textwidth]{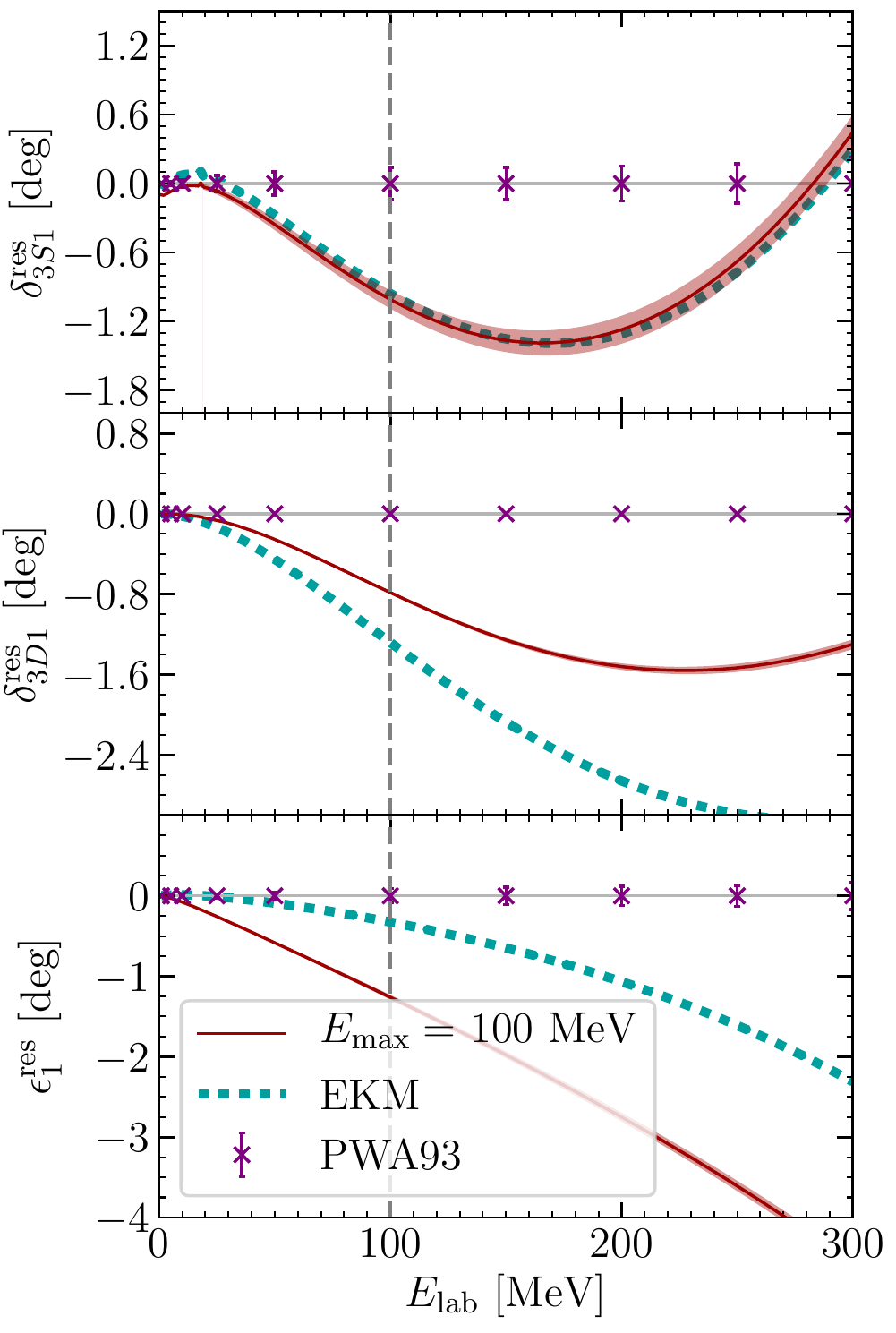}
	\caption{ 
   (a) Posterior for the NLO fit in the \threeSone\ channel 
  to the PWA93 $np$ phase shifts with $\Emax=100\,$MeV.
  The grey solid lines and square show the corresponding EKM values.
  (b) The red band is the corresponding propagated
  prediction for phase shifts and the mixing angle. The blue line is EKM's result.
	\label{fig:phaseshift_post_3S1_NLO_Emax_100MeV}}
\end{figure*}

Another case in two dimensions is shown in figure~\subref*{fig:phaseshift_post_1S0_NLO_Emax_100MeV}:
there we present the posteriors for the two \oneSzero\  LECs $\widetilde{C}_{1S0}$ and $C_{1S0}$ at NLO ($\kord=2$). 
In contrast to figure~\subref*{fig:phaseshift_post_3P0_N3LO_Emax_200MeV}, these
LECs are highly anti-correlated. 
They are well-constrained and 
the posterior pdf is Gaussian to a good approximation.
The values of $\widetilde{C}_{1S0}$ 
and $C_{1S0}$ we extract at the nominal $\Emax$
differ from those found by EKM by significantly more than the LEC uncertainties. 
This leads to a slightly different prediction, as
shown in the phase-shift residual comparison in
figure~\subref*{fig:phaseshift_1S0_NLO_Emax_100MeV}. 
The main difference is that the EKM prediction precisely reproduces the lowest-energy
phase shifts, while our prediction deviates slightly near $\sim1$--$5\mev$.

Finally, we turn our attention to the \threeSoneDone\ channel at NLO ($\kord=2$), where there are 
three contact LECs. The posterior pdfs for a fit to the PWA93 
up to $\Emax=100$ MeV are shown in 
figure~\ref{fig:phaseshift_post_3S1_NLO_Emax_100MeV}(a).
The two-dimensional histograms clearly display the correlation structure of the LECs,
just as in the previous two examples.
We see that, for example, $C_{3S1-3D1}$ is not very correlated with 
$C_{3S1}$, instead it is more correlated with the leading-order
LEC $\widetilde{C}_{3S1}$. The residuals for the propagated phase shifts and 
mixing angle for our fit are shown in figure~\ref{fig:phaseshift_post_3S1_NLO_Emax_100MeV}(b).
Our results differ somewhat from the EKM ones, particularly
in the case of the mixing angle and $\delta_{3D1}$. A fit to $\Emax~=~200\mev$
(not shown) produces very similar results to EKM
for $\delta_{3S1}$ and $\delta_{3D1}$, but the behaviour of $\epsilon_1$ above $100\mev$ remains rather different. 

\begin{figure*}[tbh]
\centering
        \includegraphics[width=0.98\textwidth]{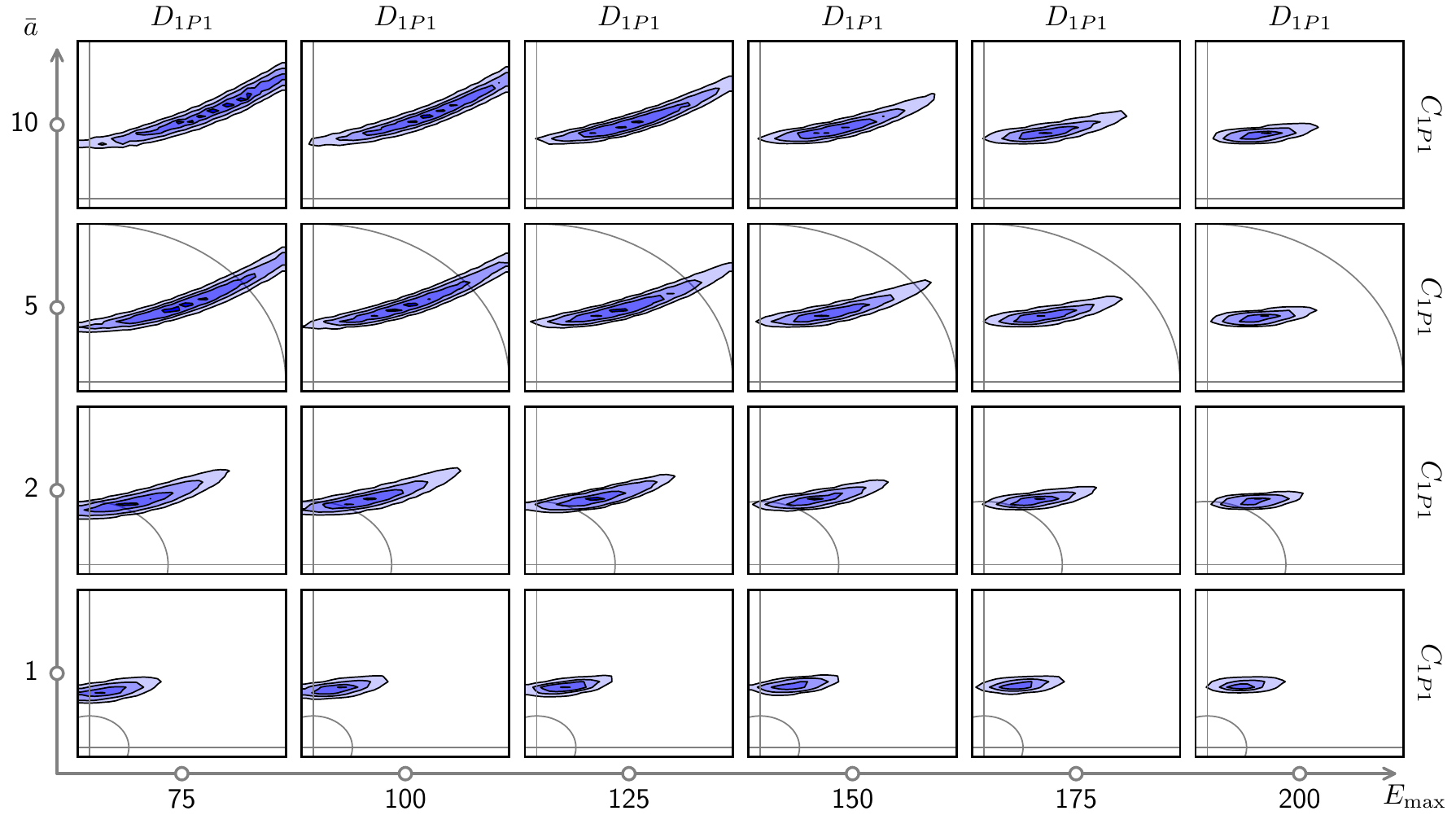}
        \caption{Posterior plots for a subset of the \NNNLO\ fits in the \onePone\ channel
          to PWA93 $np$ phase shifts from table~\ref{tab:prior-compare-LECs-1P1-N3LO} with $E_{\mathrm{max}}$ 
          ranging from 75 to 200\,MeV and $\abar$ equal to 1, 2, 5, and 10. 
          All plots are on the same scale, with the origin at the intersection of the vertical and horizontal lines
          and the circular arcs centered at the origin with a radius given by the $\abar$ for that plot. 
         \label{fig:histogram_grid_abar_vs_Emax}}
\end{figure*}

So far, all of the posterior pdfs we have considered are tightly determined
by the PWA93 phase shifts. In these cases, the precise data constrain the 
likelihood so much that the naturalness prior on the LECs with $\abar=5$ is
largely irrelevant. However, if we decrease the amount and energy range of data used to constrain
the LECs, the naturalness prior can have a strong effect on the final posterior
pdf and quoted LEC values. 
In table~\ref{tab:prior-compare-LECs-1P1-N3LO} we give an example of this
effect for the \onePone\ channel at \NNNLO\ ($\kord=4$).
We present the two LEC values with projected widths from fits to the 
partial-wave cross sections, as we will do in section~\ref{sec:Emaxplots},
noting the significantly larger uncertainties adopted (see section~\ref{sec:furtherdetails}).
We consider different values of $\Emax$ and different widths $\abar$ of the naturalness prior.
(Note: in~\cite{Wesolowski:2015fqa} the impact of $\abar$ was visualized using a diagnostic called
an ``$\abar$ relaxation plot'' rather than with tabular data.) 
We supplement the table with posterior plots in figure~\ref{fig:histogram_grid_abar_vs_Emax} for a
subset of the $\Emax$ and $\abar$ combinations.

\begin{table}[t!]\centering
\caption{Table of LEC values at different $\Emax$ in the \onePone\ channel
using the potential at \NNNLO, where there are two contact LECs. 
The fit is to the partial wave cross section with the larger uncertainties used in
section~\ref{sec:Emaxplots}.
No  variance was added to account for the EFT truncation uncertainty ($\Delta\genobsvecth = 0$).
We compare the median LEC values and their central 68\% credible intervals extracted using four different widths $\abar$ for
the Gaussian naturalness prior for a range of $\Emax$ values.}
\vspace{3mm}
\renewcommand{\crule}[1]{\multispan{#1}{\hrulefill}}
\lineup
\small
\begin{tabular}{@{}Sr*{11}{Sc}}\br
& \multicolumn{5}{c}{$C_{1P1}$} && \multicolumn{5}{c}{$D_{1P1}$} \\
\ns \multicolumn{1}{c}{\!\!\Emax\!\!} & \crule{5} & & \crule{5} \\
\relax [MeV]\!\!\!\! & $\abar=1$ & $\abar=2$ & $\abar=5$ & $\abar=10$ & $\abar=20$ &  & $\abar=1$ & $\abar=2$ & $\abar=5$ & $\abar=10$ & $\abar = 20$ \\
\mr
25 & $1.5_{-0.2}^{+0.2}$ & $1.7_{-0.3}^{+0.5}$ & $2.4_{-0.8}^{+2.2}$ & $5.2_{-3.3}^{+6.1}$ & $12_{-8}^{+14}$ &  & $0.0_{-0.9}^{+0.9}$ & $0.5_{-1.7}^{+1.6}$ & $2.5_{-3.0}^{+3.2}$ & $6.3_{-5.0}^{+4.8}$ & $11_{-7}^{+7}$ \\
50 & $1.7_{-0.2}^{+0.2}$ & $1.9_{-0.3}^{+0.6}$ & $2.8_{-1.0}^{+2.1}$ & $5.4_{-3.0}^{+5.7}$ & $12_{-8}^{+14}$ &  & $0.1_{-0.8}^{+0.8}$ & $0.9_{-1.4}^{+1.5}$ & $3.2_{-2.6}^{+2.9}$ & $6.6_{-4.2}^{+4.5}$ & $11_{-7}^{+7}$ \\
75 & $1.8_{-0.2}^{+0.2}$ & $2.0_{-0.3}^{+0.5}$ & $3.0_{-1.0}^{+2.0}$ & $4.6_{-2.2}^{+5.0}$ & $8.5_{-5.3}^{+12}$ &  & $0.4_{-0.8}^{+0.8}$ & $1.3_{-1.3}^{+1.3}$ & $3.5_{-2.2}^{+2.7}$ & $5.8_{-3.4}^{+4.4}$ & $9.4_{-5.6}^{+6.9}$ \\
100 & $1.9_{-0.2}^{+0.2}$ & $2.1_{-0.3}^{+0.5}$ & $2.7_{-0.7}^{+1.5}$ & $3.6_{-1.3}^{+3.2}$ & $4.4_{-2.0}^{+7.0}$ &  & $0.6_{-0.7}^{+0.7}$ & $1.5_{-1.1}^{+1.2}$ & $3.2_{-1.8}^{+2.3}$ & $4.6_{-2.6}^{+3.6}$ & $5.8_{-3.3}^{+6.0}$ \\
125 & $1.9_{-0.1}^{+0.2}$ & $2.1_{-0.2}^{+0.4}$ & $2.4_{-0.4}^{+0.9}$ & $2.6_{-0.6}^{+1.3}$ & $2.8_{-0.7}^{+1.8}$ &  & $0.8_{-0.7}^{+0.7}$ & $1.6_{-1.0}^{+1.1}$ & $2.6_{-1.4}^{+1.8}$ & $3.1_{-1.6}^{+2.3}$ & $3.3_{-1.8}^{+2.8}$ \\
150 & $2.0_{-0.1}^{+0.2}$ & $2.1_{-0.2}^{+0.3}$ & $2.2_{-0.3}^{+0.5}$ & $2.3_{-0.3}^{+0.6}$ & $2.3_{-0.3}^{+0.6}$ &  & $0.9_{-0.6}^{+0.6}$ & $1.5_{-0.8}^{+1.0}$ & $2.1_{-1.1}^{+1.3}$ & $2.2_{-1.1}^{+1.5}$ & $2.3_{-1.2}^{+1.5}$ \\
175 & $2.0_{-0.1}^{+0.1}$ & $2.1_{-0.1}^{+0.2}$ & $2.1_{-0.2}^{+0.3}$ & $2.1_{-0.2}^{+0.3}$ & $2.1_{-0.2}^{+0.3}$ &  & $0.9_{-0.5}^{+0.6}$ & $1.4_{-0.7}^{+0.8}$ & $1.7_{-0.8}^{+1.0}$ & $1.8_{-0.8}^{+1.0}$ & $1.7_{-0.8}^{+1.1}$ \\
200 & $2.0_{-0.1}^{+0.1}$ & $2.0_{-0.1}^{+0.1}$ & $2.0_{-0.1}^{+0.2}$ & $2.0_{-0.1}^{+0.2}$ & $2.0_{-0.1}^{+0.2}$ &  & $0.9_{-0.5}^{+0.5}$ & $1.2_{-0.6}^{+0.7}$ & $1.4_{-0.6}^{+0.7}$ & $1.4_{-0.7}^{+0.8}$ & $1.4_{-0.7}^{+0.8}$ \\
\br
\end{tabular}
\label{tab:prior-compare-LECs-1P1-N3LO}
\end{table}

The choice of $\abar=1$ is the most restrictive prior constraint in 
table~\ref{tab:prior-compare-LECs-1P1-N3LO}. For this $\abar$ and low values of $\Emax$, which 
correspond to small amounts of constraining data and a limited energy range, 
the estimate of the \NNNLO\ contact term $D_{1P1}$ is seen to be driven by the choice of the prior because it
is close to zero with a width of unity for the lowest values of $\Emax$. This
is an example of ``returning the prior'', where the likelihood so poorly constrains
the parameter value that the prior is the dominant input rather than the data. 
The consequence is that the two LECs are prevented from playing off each other, leading
to a more reliable estimate of $C_{1P1}$.
This constraint is released as the prior width is relaxed from $\abar=1$ to $\abar=20$,
leading to values of both $C_{1P1}$ and $D_{1P1}$ for low values of $\Emax$ showing 
strong sensitivity to $\Emax$.
This sensitivity systematically decreases as $\Emax$ increases, and once
$\Emax\approx150\mev$, the LEC estimates are statistically the same for $\abar \gtrsim 2$, while the remaining deviation of $D_{1P1}$ for $\abar=1$ implies it is too restrictive. 
These features are also seen in figure~\ref{fig:histogram_grid_abar_vs_Emax}, but with the induced correlations
between the LECs now evident at smaller $\Emax$ and larger $\abar$.
Because this example is at \NNNLO\ (and no truncation error has been
included), we expect to need more data at higher energies to 
constrain both LECs. 
For the examples in this paper
we generally have enough data that the prior sensitivity is small.

Figures \ref{fig:both_phaseshift_1P1_N2LO_Emax_100MeV} through \ref{fig:phaseshift_post_3S1_NLO_Emax_100MeV} provide specific examples of the ways projected posterior pdfs can
display information on the joint distribution(s) of LECs. Note that each pdf is ultimately close to normal, but we had to verify this explicitly.
For some cases when we obtain the LECs from the PWA93 phase
shifts using the nominal $\Emax$ value quoted in \cite{Epelbaum:2014efa,Epelbaum:2014sza} our results for the LECs---and hence our predictions for phase shifts---are not consistent with EKM's. 
However, sometimes by adjusting $\Emax$ we can find a dataset for which our results agree with theirs.
In examples
such as in figure~\subref*{fig:phaseshift_1S0_NLO_Emax_100MeV}, we find that the obtained
phase shifts reproduce the lowest-energy phase shifts less precisely than EKM, although
they are within our stated 68\% credible intervals for the prediction. This shows the
importance of considering uncertainty propagation in predictions when comparing to data.
We return to the sensitivity to $\Emax$ in section~\ref{sec:Emaxplots}. 
But first we consider the case of $s$-waves at \NNNLO, where the projected posteriors are
anomalous compared to those in this section, possibly signalling a physics issue to uncover.

\subsection{Analysis of \texorpdfstring{$s$}{s}-wave contact LEC extraction at higher order} \label{sec:swave-case}

In the \oneSzero\  channel at \NNNLO\ ($\kord = 4$) there are four \NN\ contacts
present: $\widetilde{C}^{np}_{1S0}$, $C_{1S0}$, 
$D^1_{1S0}$, and $D^2_{1S0}$. There are several
different maxima for the \NNNLO\ [and the \NNNNLO\ ($\kord=5$)] fit. 
The general analysis of a posterior
pdf involves identifying all of the modes and their structure. Multiple
modes often arise in nonlinear parameter-estimation problems, and 
it may be possible to further constrain the posterior pdf by including
more prior information, such as the priors used at higher 
orders by EKM~\cite{Epelbaum:2014efa}. 
But choosing modes in general is difficult 
and finding all of them by MCMC sampling typically requires starting
the sampling in different parts of the parameter space and may be intractable.
A particularly complicated structure may be a signal of a problem with the physics,
as we find here.

\begin{figure*}[tbh]
\centering
        \includegraphics[width=0.9\textwidth]{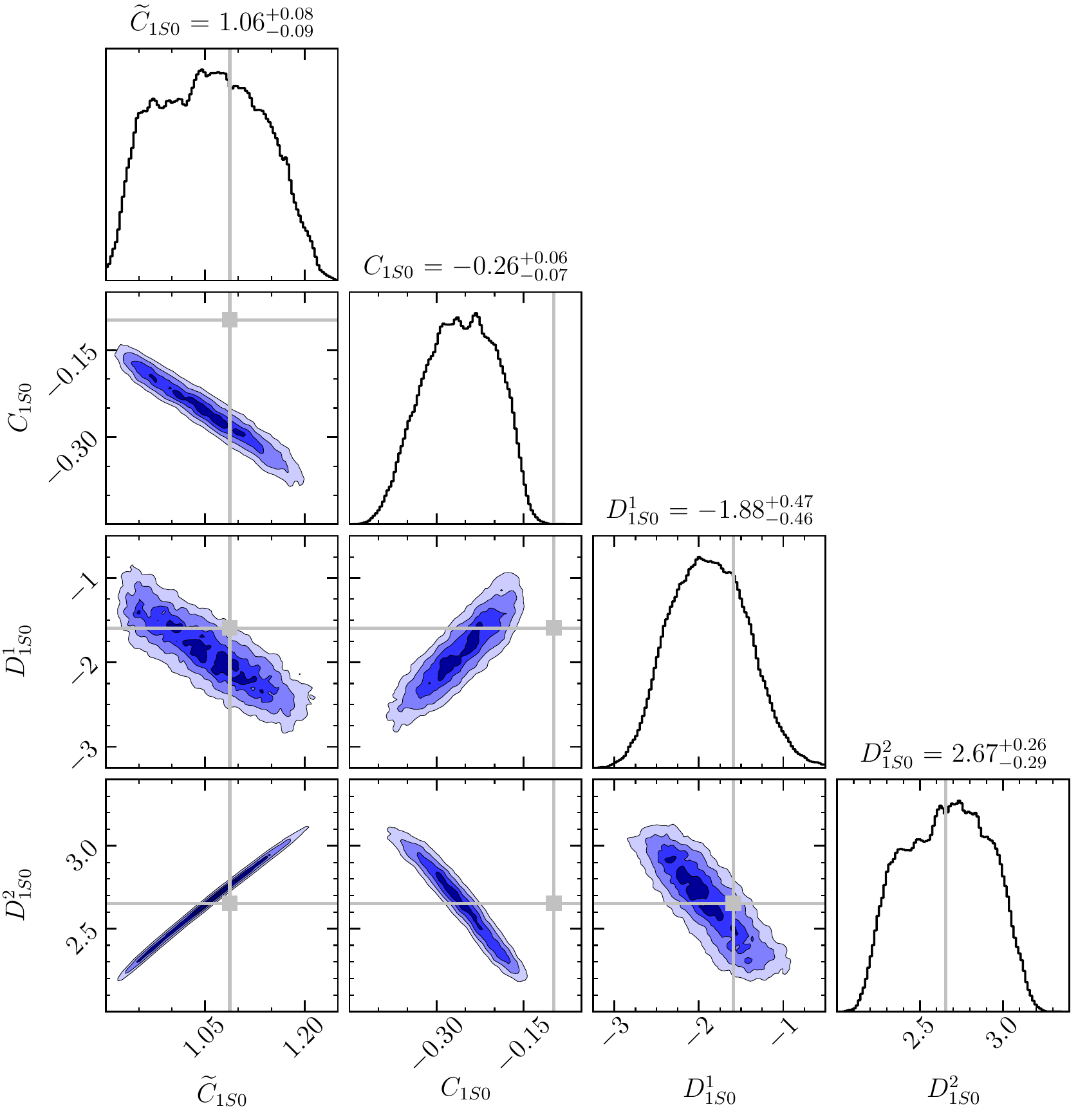}
        \caption{ Posterior plot for the \NNNLO\ fit in the \oneSzero\  channel 
          to PWA93 $np$ phase shifts with $E_{\mathrm{max}}=200\,$MeV. 
          The corresponding EKM values
          are $\widetilde{C}_{1S0} = 1.09$, $C_{1S0} = -0.10$, $D^1_{1S0}=-1.59$,
          and $D^2_{1S0}=2.65$.
         \label{fig:phaseshift_post_1S0_N3LO_Emax_200MeV}}
\end{figure*}

\begin{figure*}[ptbh]
    \centerline{\includegraphics[width=0.5\textwidth]{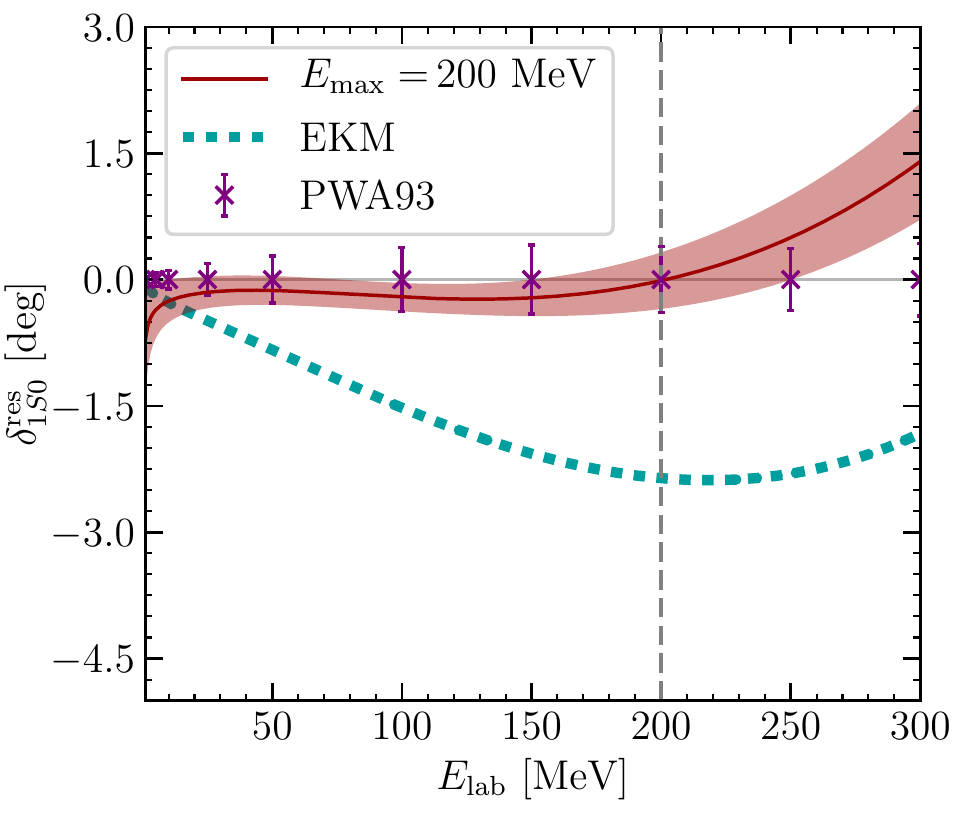}}
  \caption{The red band shows the phase shifts in the \oneSzero\ channel at \NNNLO\ propagated from the pdf in  figure~\ref{fig:phaseshift_post_1S0_N3LO_Emax_200MeV}. The teal line is the EKM result.
  \label{fig:phaseshift_1S0_N3LO_Emax_200MeV}}
\end{figure*}

Figure~\ref{fig:phaseshift_post_1S0_N3LO_Emax_200MeV} shows the result
of estimating the four LECs at \NNNLO\ from the PWA93 phase shifts
up to $\Emax=200\,\text{MeV}$. 
(We use the mode that gives
phase shifts closest to the PWA93.)
The posterior pdf deviates from normality---unlike those we saw
in section~\ref{sec:post-information}: it has concentrated pdfs on the diagonal and significant correlations between
the parameters. The corresponding propagated phase shifts are shown in 
figure~\ref{fig:phaseshift_1S0_N3LO_Emax_200MeV}. (Note that a covariance matrix
method was used even though the pdf is no longer Gaussian. This approximation
is adequate for our purposes here, because
the uncertainty is small.)
The fit is quite good visually and has an acceptable $\chi^2/\mathit{dof}$, even though the joint
pdfs of the parameters at this order have a number of peculiar features.

\begin{figure*}[ptbh]
	\centering
	\includegraphics[width=0.60\textwidth]{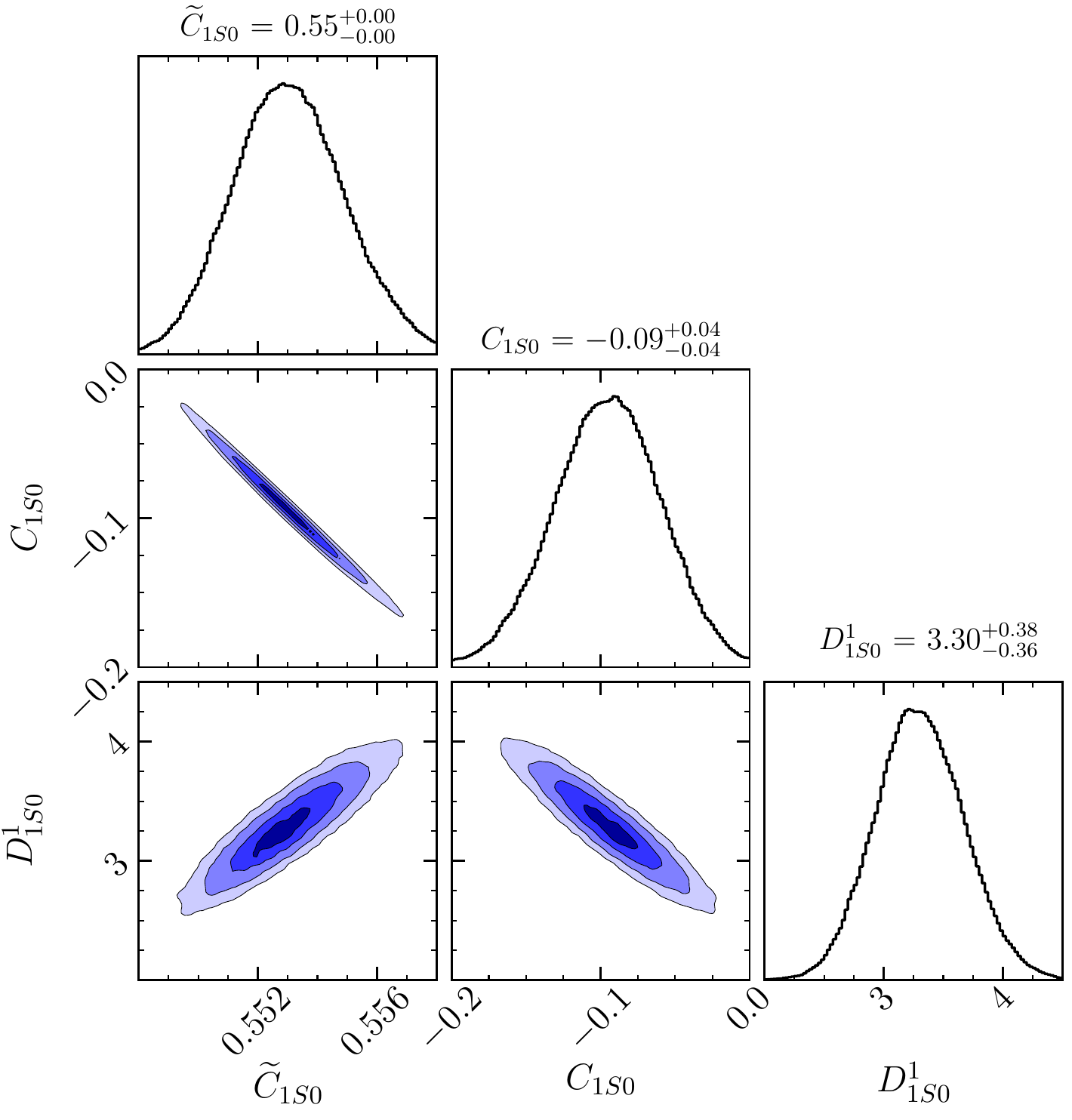}~%
	\raisebox{4cm}{\includegraphics[width=0.37\textwidth]{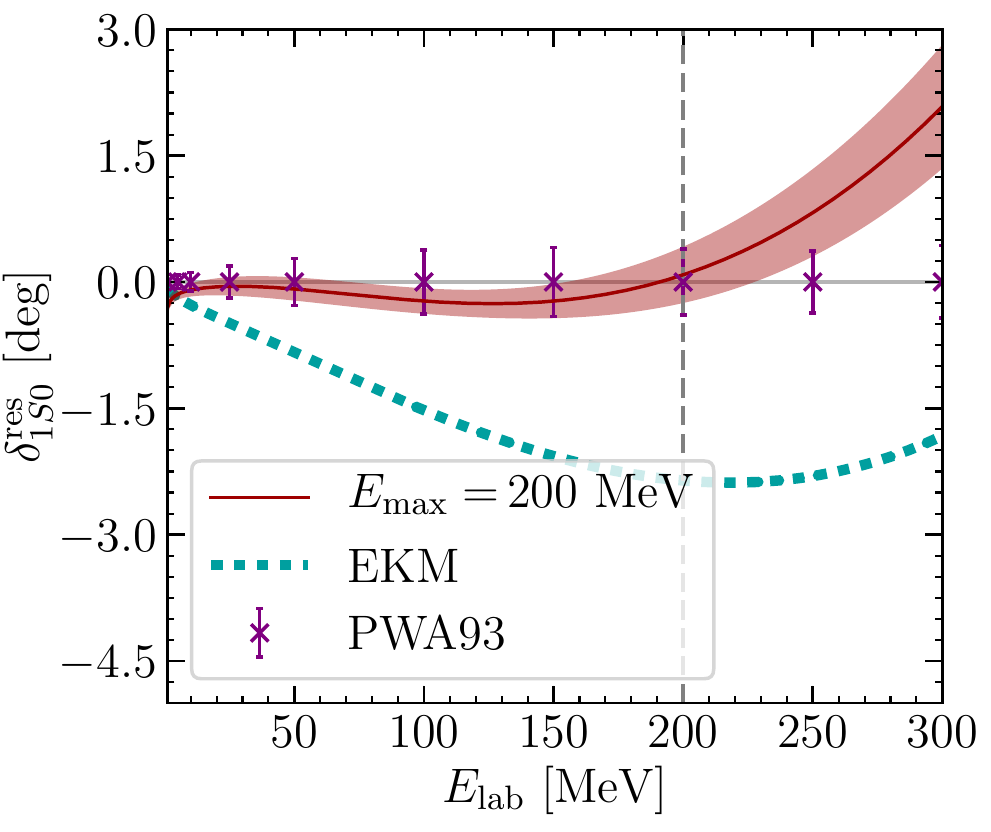}}
	\caption{
	(a) Posterior plot for the \NNNLO\ fit in the \oneSzero\  channel 
	to PWA93 $np$ phase shifts with $E_{\mathrm{max}}=200\,$MeV.
	The LEC $D^2_{1S0}$ was fixed
	to zero for this fit. 
	We show the resulting value of $D^1_{1S0}$,
	which fully determines the value of $D^2_{1S0}$. 
	(b) The red band shows the resulting 
  prediction for the \oneSzero\ phase shifts. The blue line is the EKM result.
	\label{fig:phaseshift_post_1S0f2_N3LO_Emax_200MeV}%
	}
\end{figure*}

Those oddities of the \NNNLO\ pdf are not
present in, e.g.\ the NLO pdf that results in Figs.~\subref*{fig:phaseshift_post_1S0_NLO_Emax_100MeV}
and~\subref*{fig:phaseshift_1S0_NLO_Emax_100MeV}. This leads us to examine their 
origin~\cite{Wesolowski:2016int}. 
We find they are symptomatic of a parameter degeneracy in 
the \NN\ potential at \NNNLO. The short-range part of the 
\NN\ potential matrix element in the \oneSzero\  channel 
at \NNNLO\
can be rearranged as
%
\beq
  \begin{split}
  \langle {}^1S_0 | V_{NN} | {}^1S_0 \rangle & = D^1_{1S0}\, p^2 p^{\prime2} + 
  D^2_{1S0}\, (p^4 + p^{\prime4} )\\
  & = \frac{1}{4} (D^1_{1S0} + 2 D^2_{1S0})(p^2 + p^{\prime2} )^2 
      - \frac{1}{4} (D^1_{1S0} - 2 D^2_{1S0})(p^2 - p^{\prime2} )^2 \\
   & = (D^1_{1S0} + 2 D^2_{1S0}) \, p^2 p^{\prime2} + D^2_{1S0} (p^2 - p^{\prime2} )^2 \;,
  \end{split}
  \label{eq:swave-rewrite}
\eeq
where the second term on the right side of both the second and third line of  \eqref{eq:swave-rewrite} vanishes
on-shell. (Similar rearrangements are also possible for the contact pieces of the ${}^3S_1$-to-${}^3S_1$
\NN\ matrix element and the mixing term.) As shown in \cite{Reinert:2017usi}, the operator  
$(p^2 - p^{\prime2})^2$ can be 
transformed into higher-order and/or higher-body operators using unitary transformations. 
(These are equivalent in this case to 
redefinitions of the nucleon field in the EFT Lagrangian, see \cite{Furnstahl:2000we}.)  
This implies that the parameter $D_{1S0}^2$ may not affect
the \NN\ data at \NNNLO\ accuracy:  we anticipate a parameter 
redundancy in any fit at this order
that includes both of the short-range $Q^4$ operators in the first line of  (\ref{eq:swave-rewrite}).

In the EFT of the \NN\ system with only short-range interactions, 
the so-called ``pionless EFT''~\cite{Kaplan:1998we,vanKolck:1998bw,Birse:1998dk,Bedaque:2002mn}, the equivalence of on-shell matrix elements obtained
with different values of $D^2_{1S0}$ is guaranteed. 
The operators $(p^2 + p^{\prime \, 2})^2$ and $(p^2 - p^{\prime \, 2})^2$ both enter as \NNLO\ perturbations in the
pionless-EFT \NN\ amplitude.  
As we show in \ref{app:redundantpionless}, in this 
case the second term in  (\ref{eq:swave-rewrite}) can 
be exactly absorbed into lower-order contributions to the
$T$-matrix. This is most easily seen in dimensional regularization~\cite{Kaplan:1998sz},
but is true in any other regularization scheme too~\cite{vanKolck:1998bw}.

In \eft\ the situation is not as clear, because these operators are 
treated non-perturbatively, and they mix with long-range pion physics. 
We therefore now explore the expected parameter degeneracy numerically. 
Following \cite{Reinert:2017usi}, we fix the off-shell combination 
$D^2_{1S0}$ to zero. In this case, we discover only a single
mode using MCMC sampling.
The description of the data is shown in figure~\ref{fig:phaseshift_post_1S0f2_N3LO_Emax_200MeV}(b);
as observed in \cite{Reinert:2017usi} it is just as good as
that in figure~\ref{fig:phaseshift_1S0_N3LO_Emax_200MeV}, which was
generated from the posterior that includes an additional parameter $D^2_{1S0}$ associated 
with the $(p^2 - p'^2)^2$ operator.

However, once again, we find that the posterior pdf contains more information than, say, the $\chi^2$ value. 
The posterior pdf that results from setting $D^2_{1S0}=0$ is shown in 
figure~\ref{fig:phaseshift_post_1S0f2_N3LO_Emax_200MeV}(a).
It  follows a Gaussian distribution out beyond the point of $2$-$\sigma$. But, 
  (\ref{eq:swave-rewrite}) makes it clear that there are (infinitely) many equivalent
ways to eliminate the off-shell operator $p^2 - p'^2$. If we instead adopt the choice implied by
the second line of  (\ref{eq:swave-rewrite}) then the description of data remains good, but the 
posterior does not follow a Gaussian distribution. We hypothesize that this is because the 
operator $(p^2 + \ppsq)^2$, when iterated, generates divergences proportional to a
much higher power of the cutoff than does iteration of the operator $p^2 p'^2$. In pionless EFT
$(p^2 + \ppsq)^2$ produces divergences proportional to $\Lambda^9$, while $p^2 p'^2$ only yields
quintic divergences. And indeed, \cite{Reinert:2017usi} opted to set $D^2_{1S0}$ to zero, 
that is, to use the second of the two rewritings in  (\ref{eq:swave-rewrite}) rather than the first,
because it produces a softer, and therefore more perturbative, \NN\ potential. The $p^2 p'^2$ choice may also be advantageous because 
 $p^2 p'^2$ is exactly the operator that---together with the short-range operators included at NLO---ensures a closed basis of short-range operators,  $\{1,p^2,p'^2,p^2 p'^2\}$, under iteration. 

This case study shows how correlations between \eft\ parameters in the same channel, at the same order, 
together with the non-Gaussianity
evident in figure~\ref{fig:phaseshift_post_1S0_N3LO_Emax_200MeV}, indicate an 
issue with the parameter estimation  in the \oneSzero\  channel at \NNNLO\ of the chiral expansion. 
Using the rearrangement in  \eqref{eq:swave-rewrite}
we---and \cite{Reinert:2017usi}---traced this issue to the fact that one of the \NNNLO\ short-range operators
can be unitarily transformed into higher-order/higher-body operators. The operator  $(p^2 - \ppsq)^2$ therefore
does not contribute to the \NN\ problem at fourth (or fifth) order, and  its LEC $D^2_{1S0}$ can be set to zero 
without affecting the quality of the fit. The posterior pdf that results when $D^2_{1S0}$ is set to zero
shows that using only the three remaining \oneSzero\  parameters at \NNNLO\ produces an equally good description
of the data. Furthermore, the posterior pdfs for two different choices of \NNNLO\ operator bases 
show that one basis choice---the softer one---produces a posterior with more
regular, Gaussian behaviour.
Posterior pdfs are the result of the given information: in this case the chosen data and
the behaviour of the input potential produce distinct features in the posterior pdf that lead to 
physical insight.

\section{Case study 2: LEC stability with maximum energy for fit} \label{sec:Emaxplots}

A question that can bedevil EFT parameter estimation is that of the
energy range of the data that should be used. In general the LECs will
be determined more precisely as more data are considered. However, if those
data are at higher energies this precision may be spurious, as more higher-order
terms in the EFT start to contribute to the parameter estimation in that region.
EKM chose to deal with this question by increasing the $\Emax$ for their fit
as they considered higher orders in the EFT expansion, but, as noted in 
section~\ref{sec:post-information}, the fit LECs are sensitive to how this is done. 
This does not imply that the quality of the fit as measured by the $\chi^2$ is strongly sensitive.
But a properly formulated EFT, with a specified renormalization scale and scheme, should not have LECs
that depend on the data used to determine them.
Here we apply our Bayesian framework to this problem, and show that proper uncertainty quantification
produces LECs that do not depend on the $\Emax$ of the fit, to within statistical uncertainty. 

\begin{figure*}[thb]
\centering
   \includegraphics[width=0.48\textwidth]{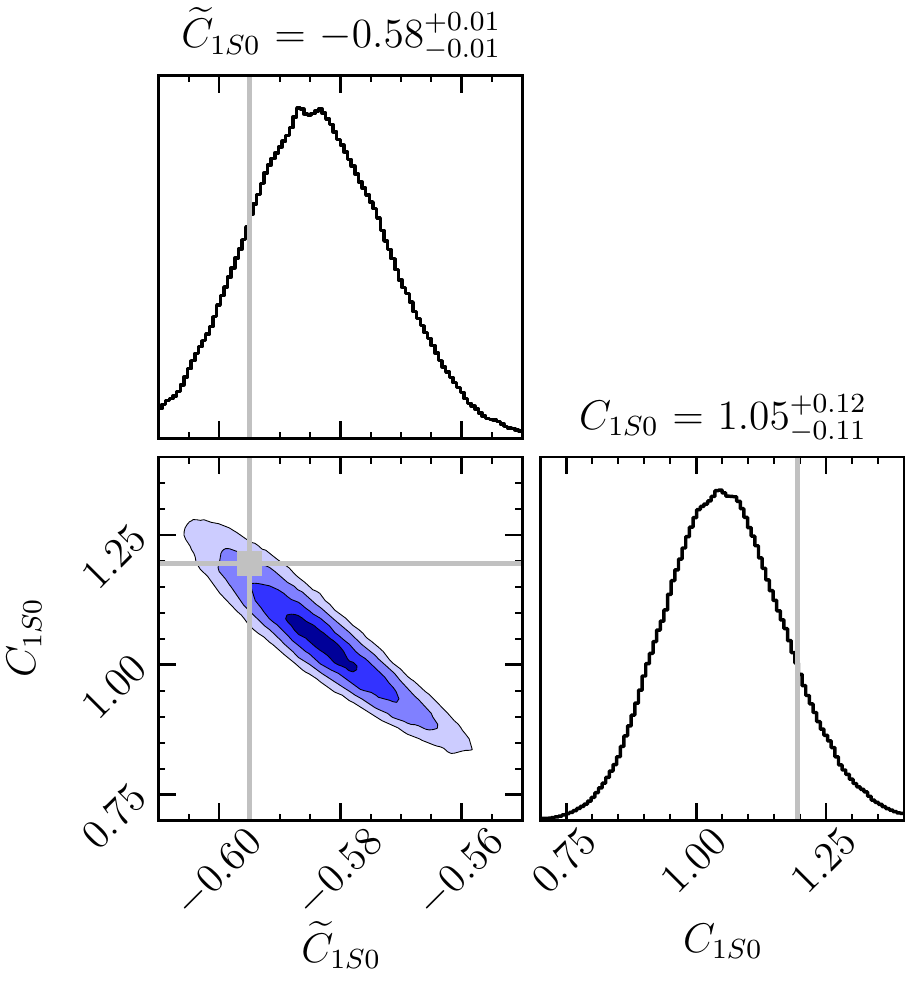}%
   \includegraphics[width=0.48\textwidth]{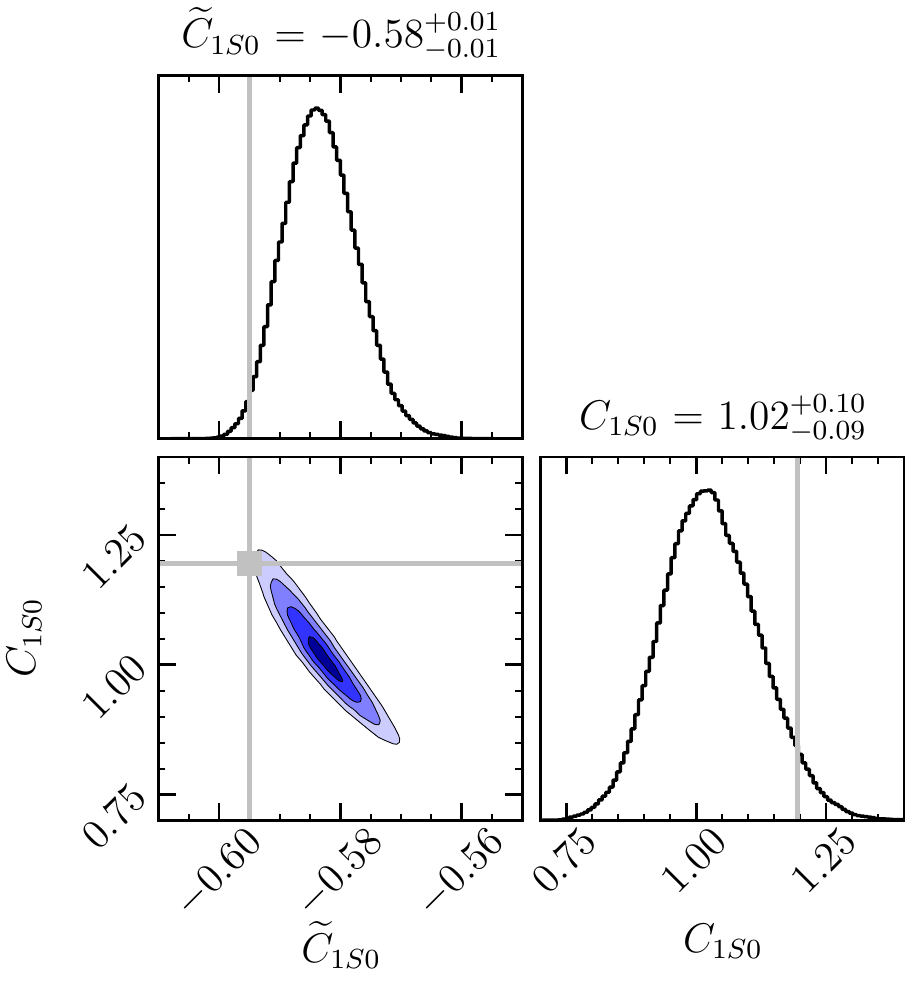}
   \caption{Posteriors for the NLO fit to the $np$ partial-wave cross sections in the \oneSzero\  channel
     with $E_{\mathrm{max}}=100\,$MeV. Here truncation errors are accounted for by
     using equation \eqref{eq:full_posterior_for_LECs} and taking $\kmax=\kord +1=3$,
     using the (a) uncorrelated and (b) fully correlated assumptions for higher-order terms (cf. 
  \eqref{eq:covariance_case_A} and \eqref{eq:covariance_case_B}).}
   \label{fig:posteriorincltruncation}
\end{figure*}

Including truncation errors in the posterior pdf for the LECs $\LECk$ stops LECs being driven by data at energies where the EFT is taken at too 
low an order to describe those data (``underfitting''). 
Therefore in this section we include $\covarth$ in \eqref{eq:full_posterior_for_LECs}, considering models for the theory error
in which that error is both uncorrelated across different energies and constant at all energies---see
\eqref{eq:covariance_case_A} and \eqref{eq:covariance_case_B}.
In a given partial wave, we assume the expansion starts from the first physically sensible order
predicted in \eft, 
but that successive terms are still suppressed by increasing powers of $Q$ with increasing order. 
We take this starting order to be LO for $s$-waves and NLO for $p$-waves,
and take $(\genobsrefvec)_i = (\genobsvecexp)_i$. 
In figure~\ref{fig:posteriorincltruncation} we
show the \oneSzero\ posterior pdf using the partial wave cross section as data.
When performing the fit, we include both the
truncation error in the first-omitted-term approximation as well as the larger experimental error described in section \ref{sec:furtherdetails} and below.
The limiting cases  \eqref{eq:covariance_case_A} and \eqref{eq:covariance_case_B} for higher-order terms
are shown.
While the central values are very close, the posteriors differ in detail.

Although in figure~\ref{fig:posteriorincltruncation} we only considered $\kmax=\kord +1$, in general we could ensure
that we have properly 
accounted for the higher-order terms by increasing $\kmax$ until the results become 
stable with respect to it,  i.e.\ we've 
marginalized over enough higher-order terms that the full effect of the EFT truncation is seen.
Or we can simply compare $\kmax = k+1$ 
to $\kmax \rightarrow \infty$, which is what we do in this section.
For a given channel and EFT order, we examine 
the ``$\Emax$ plot" at these limits of $\kmax$~\cite{Wesolowski:2015fqa}. 
Such plots show the median and 68\% credible interval for the extracted LEC (or LECs) as a function 
of the maximum energy of the data used for the extraction. 
They show us whether an LEC ``saturates'' with $\Emax$, i.e.\ if the 
 $\Emax$ plot exhibits LEC values and widths that do not change 
significantly as data at higher energies are included in the parameter estimation. 
Note that, as shown in table~\ref{tab:prior-compare-LECs-1P1-N3LO}, the LEC
values at low $\Emax$ are affected by the prior, but in the region of interest the
results are insensitive to the width of the naturalness prior on the LECs.

We include truncation errors in the parameter estimation
by assuming the naturalness of the observable expansion at higher EFT orders.
In general, it must be verified that this assumption holds for any
particular observable, but it works well for the total cross section and various
polarization observables~\cite{Melendez:2017phj,Furnstahl:2015rha}. Here we assume that the
partial-wave cross section, given by,
\beq
  \begin{split}
  \sigmapw(\Elab) & = \frac{\pi}{k^2} (2J+1) \\ \times
      & \begin{cases}
        \sin^2 \delta_{\ell s}^J  & \text{if uncoupled;} \\
         2\sin^2\epsbar_J + \cos 2\epsbar_J\, (\sin^2\bar\delta_{J+1} + \sin^2\bar\delta_{J-1})   & \text{if coupled} \;,
      \end{cases}
  \end{split}
    \label{eq:pw_cross_sections}
\eeq
has an order-by-order EFT expansion with natural coefficients.
Following EKM~\cite{Epelbaum:2014efa}, we take the error in $\sigmapw$ to be the spread in 
the different model potential predictions of that quantity in
the \NN-online database. This produces much larger errors in the LECs than the purely ``statistical" errors 
we assigned to the \NN\ phase shifts for the parameter estimation of the previous section.

\begin{figure}[tbh]
  \centering
  \includegraphics[width=0.49\textwidth]{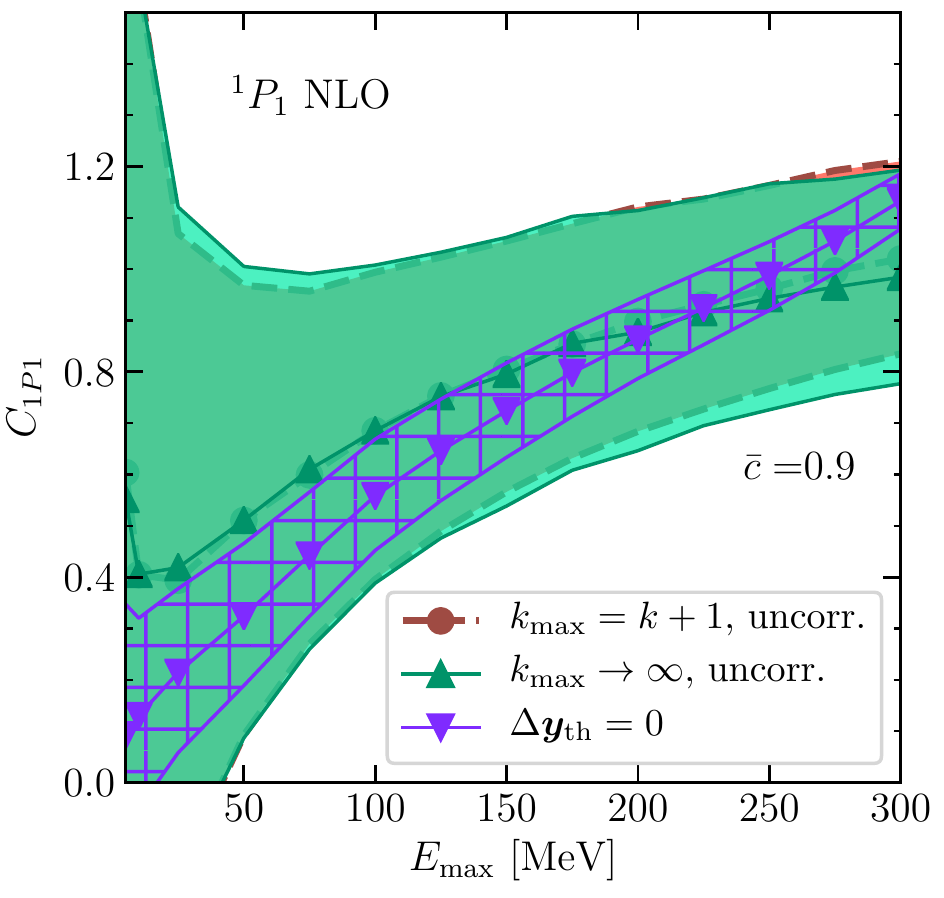}~~%
  \includegraphics[width=0.49\textwidth]{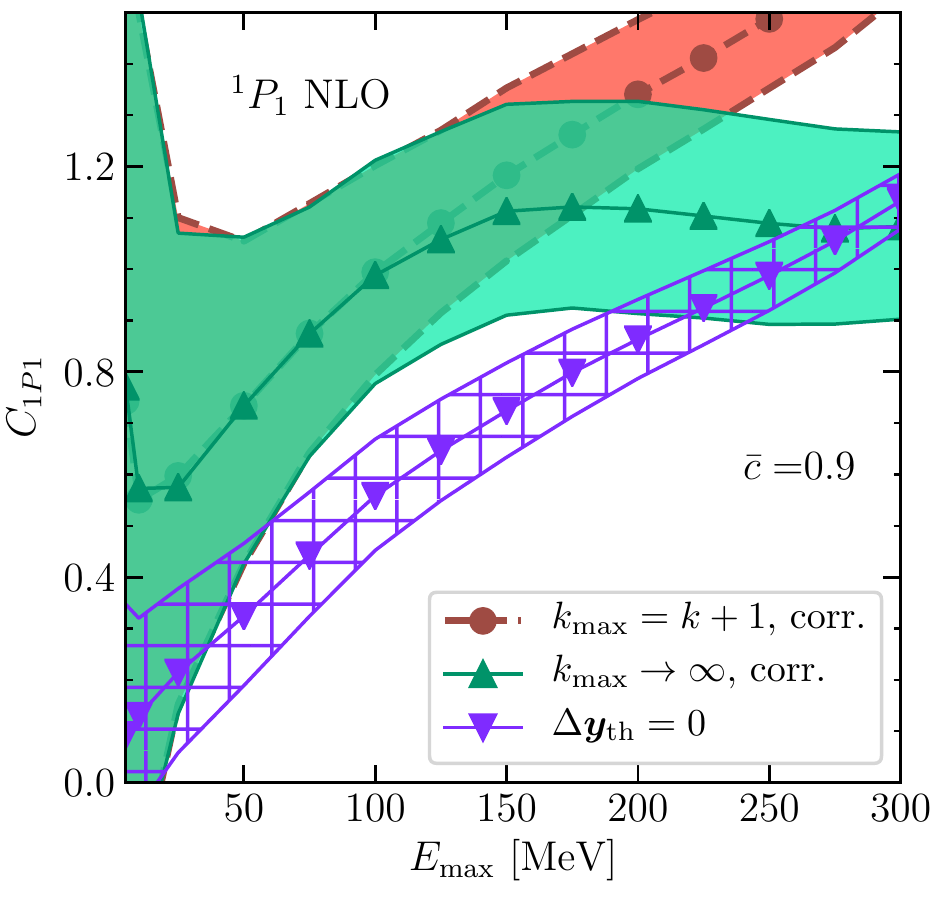}%
  \caption{$\Emax$ plots for the \NLO\ fits in the \onePone\  channel to
  the partial wave cross section.  The no-truncation-error
  results ($\Delta\genobsvecth = 0$, purple inverted triangles) are compared 
  with those adding the theoretical error using the (a) uncorrelated and (b) fully correlated assumptions as in
  \eqref{eq:covariance_case_A} and \eqref{eq:covariance_case_B}, using both the first-omitted-term
  approximation (coral circles) and the full $\kmax\to\infty$ 
  result (green triangles). The value of $\cbar = \sigmaemp=0.9$.
  \label{fig:Emax_1P1_k_1_compare_kmax_fixed_crms_0p89}}
\end{figure}

\begin{figure}[tbh]
  \centering
  \includegraphics[width=0.49\textwidth]{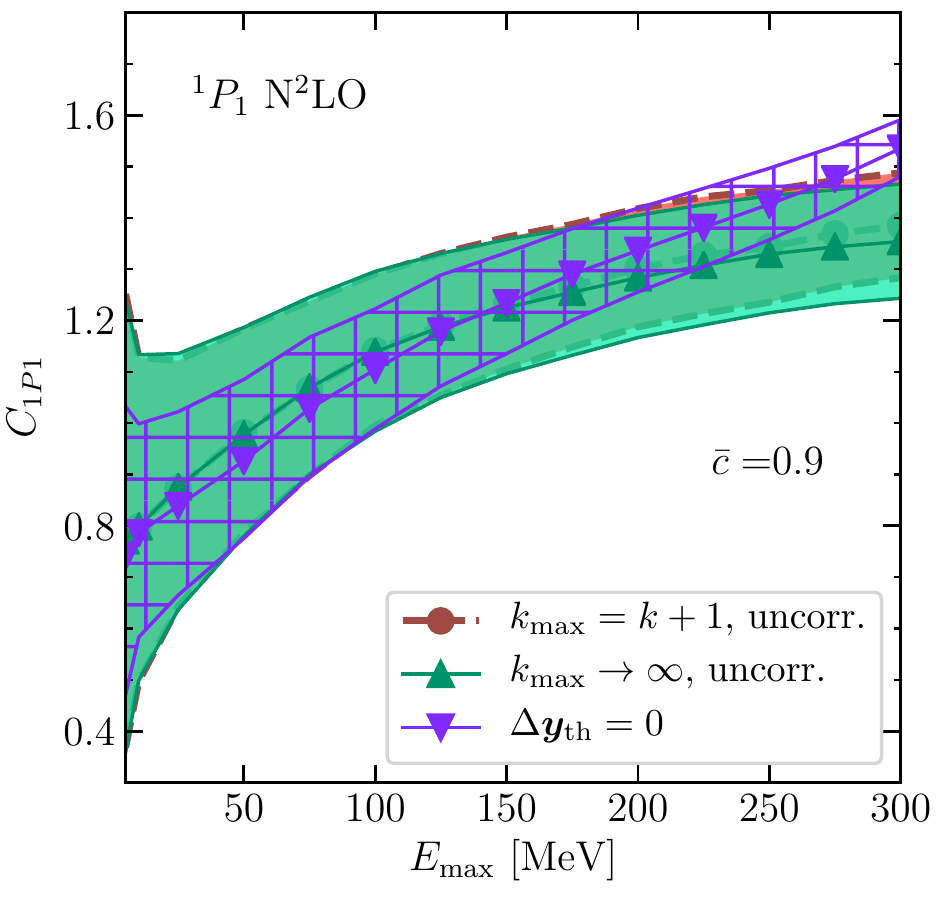}~~%
  \includegraphics[width=0.49\textwidth]{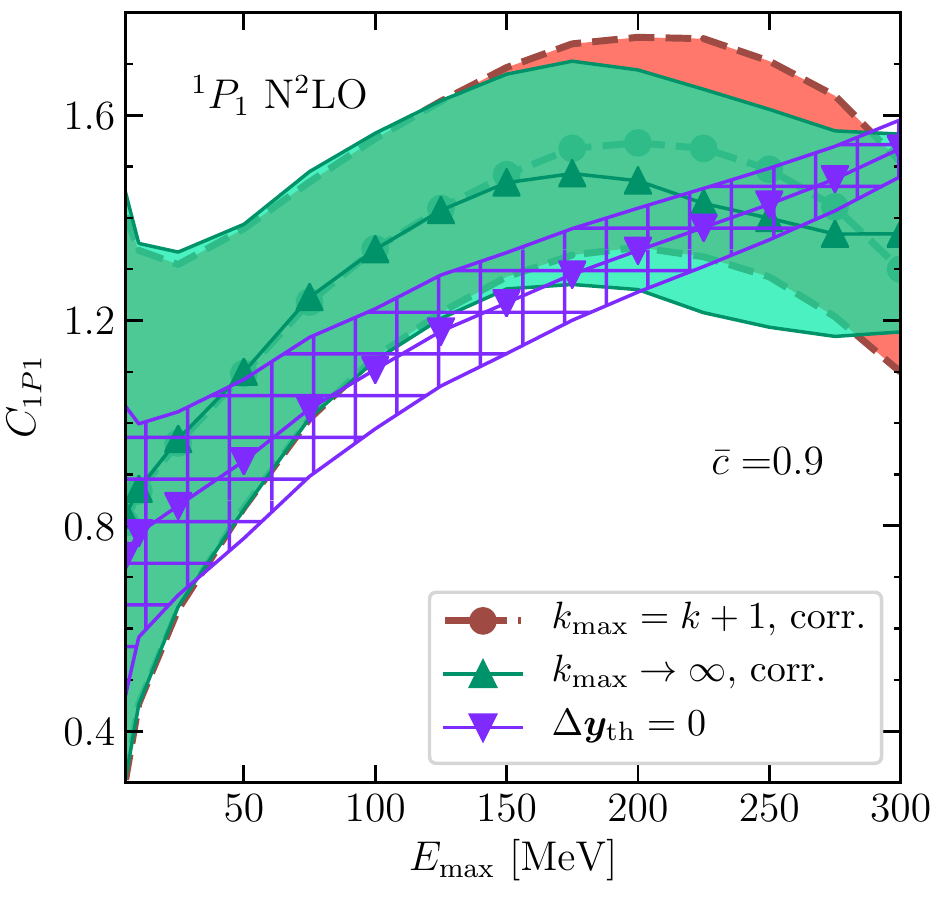}%
  \caption{Same as figure~\ref{fig:Emax_1P1_k_1_compare_kmax_fixed_crms_0p89} but for the \NNLO\ fits in the \onePone\  channel.
  \label{fig:Emax_1P1_k_2_compare_kmax_fixed_crms_0p89}}
\end{figure}

We begin by examining the $\Emax$ plots for the single \onePone\ contact LEC at \NLO\
and \NNLO.  Both panels in figure~\ref{fig:Emax_1P1_k_1_compare_kmax_fixed_crms_0p89} show the $\Delta\genobsvecth = 0$ (no truncation error) posterior maximum,
which
has a strong upward slope, changing by
about a factor of four over the full $\Emax$ range; its 68\% width is too small to make values 
over the range of $\Emax$ statistically consistent.  
Figure~\ref{fig:Emax_1P1_k_1_compare_kmax_fixed_crms_0p89} also shows 
 \NLO\ results, with (a) the uncorrelated limit
\eqref{eq:covariance_case_A} and (b) the fully correlated limit \eqref{eq:covariance_case_B}.
These are superficially rather different models for the 
theoretical discrepancy and we are at low order, so different results might be expected in panels (a) and (b). 
However, we see that in both panels the 68\% width is much wider than for $\Delta\genobsvecth = 0$.
This implies that the $\Delta\genobsvecth = 0$ uncertainty is a gross underestimate.
The bands for the uncorrelated limit of truncation error (panel (a)) are about the same for $\kmax = k+1$ and 
$\kmax\rightarrow\infty$ and yield a central value noticeably flatter in $\Emax$ than the $\Delta\genobsvecth = 0$ result. 
The LEC extraction is statistically consistent for $\Emax \geq 100$ MeV. 
In the correlated limit, $\kmax = k+1$ does not give a converged-with-$\kmax$ result, but the $\kmax\rightarrow\infty$ band flattens
out and yields results consistent with the uncorrelated case.  The differences between the 
green bands in the two panels highlight the effect of assumptions regarding the energy
dependence of the EFT truncation error and point to the need to develop 
 a statistical model for the error with a finite-energy correlation length~\cite{GPpaper:2018}.

The \NNLO\ results for \onePone\ in figure~\ref{fig:Emax_1P1_k_2_compare_kmax_fixed_crms_0p89}
exhibit the same characteristics for the $\Delta\genobsvecth = 0$ prediction, but it now only varies by about a
factor of two.  It is still the case that the width is small compared to the change in
LEC for a range in $\Emax$ of 100\,MeV or higher.
The bands on the left with the uncorrelated truncation error still overlap and are about a factor
of two smaller than in the NLO case, as expected given the higher order.
The flattening is again clear although not dramatic.   In figure~\ref{fig:Emax_1P1_k_2_compare_kmax_fixed_crms_0p89}(b) the $\kmax = k+1$ and 
$\kmax\rightarrow\infty$ bands now mostly agree, with a larger width than on the left, but with
consistent and $\Emax$-insensitive predictions, which we quote for the uncorrelated theory error as $C_{1P1}=1.4 \pm 0.1$, and $C_{1P1}=1.4 \pm 0.2$ for the 
fully-correlated theory error.

\begin{figure}[tbh]
  \centering
  \includegraphics[width=0.49\textwidth]{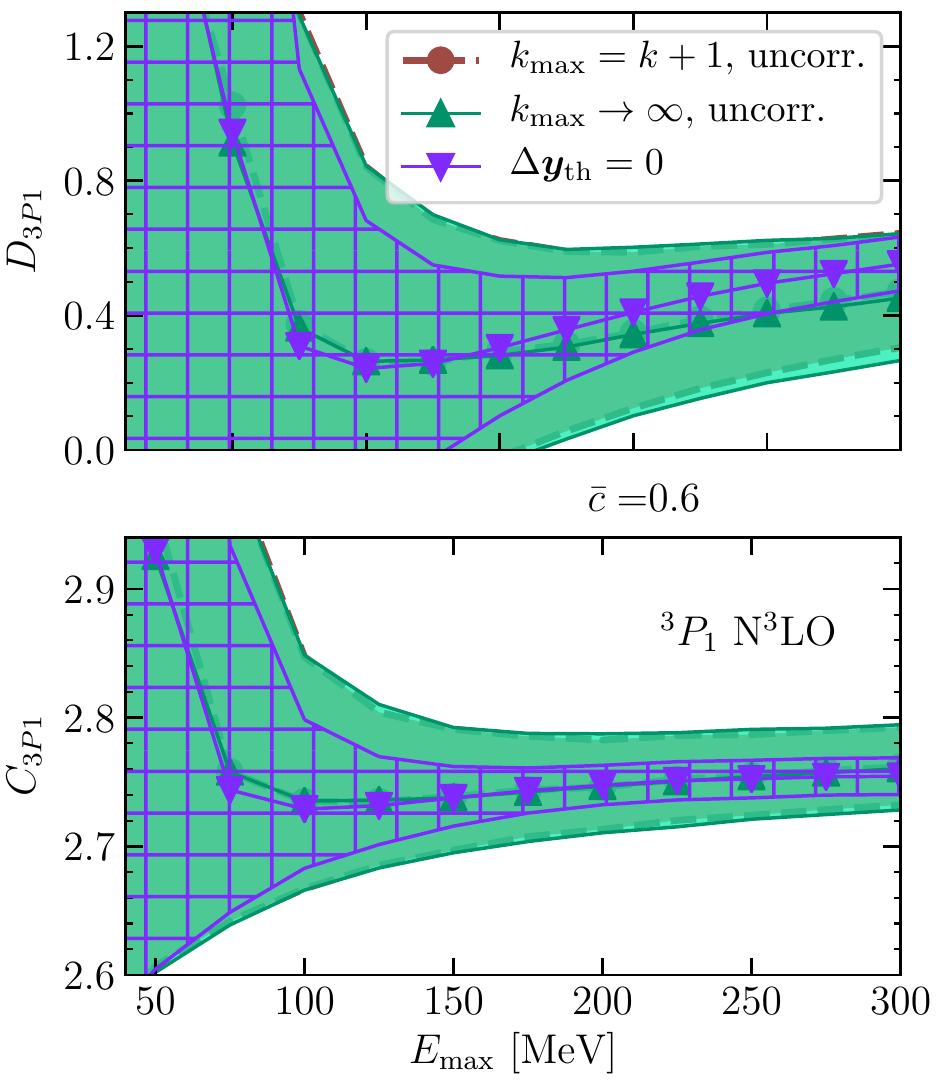}~~%
  \includegraphics[width=0.49\textwidth]{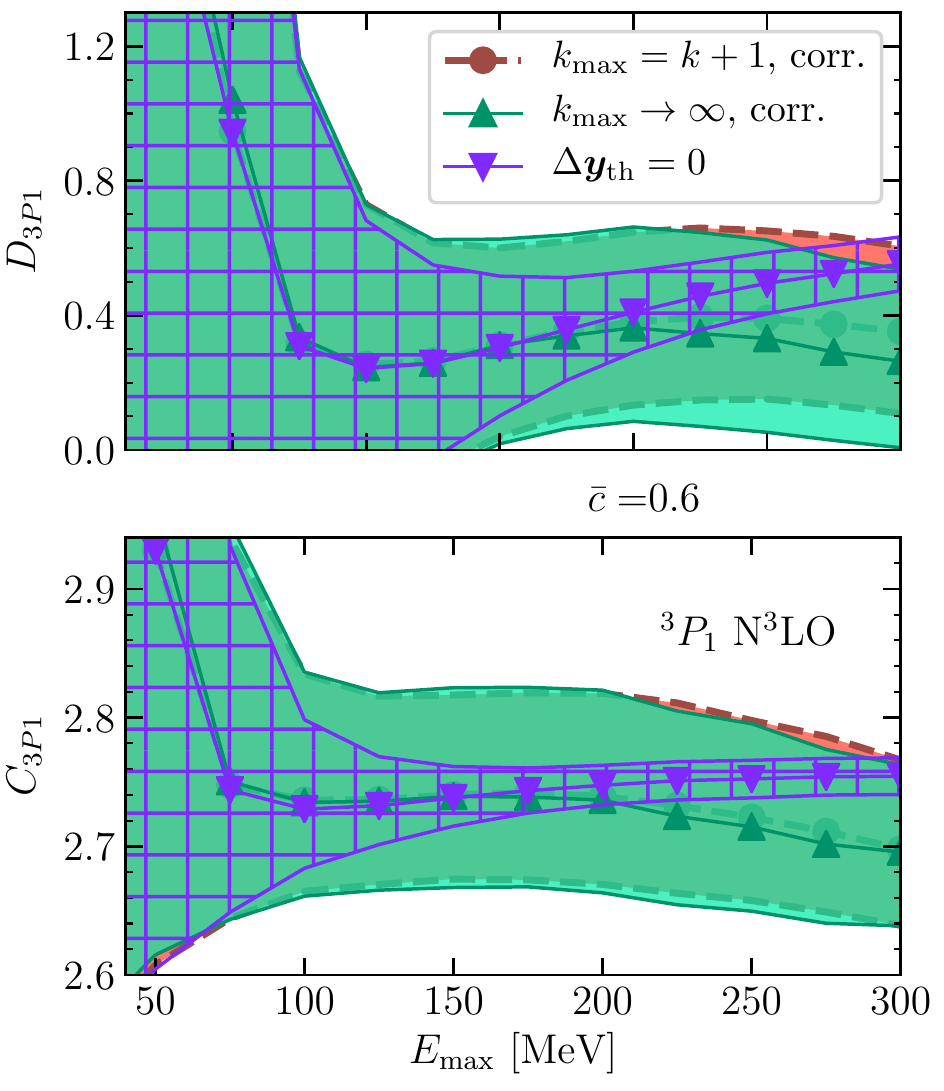}%
  \caption{$\Emax$ plots for the \NNNLO\ fits in the \threePone\ channel to the partial wave cross section.
  The no-truncation-error results ($\Delta\genobsvecth = 0$, purple inverted triangles) are compared 
  with those adding the theoretical error using the (a) uncorrelated and (b) fully correlated assumptions as in~\eqref{eq:covariance_case_A} and \eqref{eq:covariance_case_B}, using both the first-omitted-term
  approximation (coral circles) and the full $\kmax\to\infty$ 
  result (green triangles). The value of $\cbar = \sigmaemp=0.6$.
  \label{fig:Emax_3P1_k_3_compare_kmax_fixed_crms_0p6}}
\end{figure}
We expect that there will be less impact of neglected higher-order terms in
parameter estimation at higher EFT orders, i.e.\ larger values of $k$. 
So, we next examine the \threePone\  channel at \NNNLO\ ($k=4$), where there are two contact LECs.
The $\Delta\genobsvecth = 0$  results in figure~\ref{fig:Emax_3P1_k_3_compare_kmax_fixed_crms_0p6} show a flat dependence on $\Emax$
for the $C_{3P1}$ LEC but a significant slope for the subleading $D_{3P1}$ LEC.
Once again, including truncation error in the analysis produces 
LEC estimates that are less $\Emax$ dependent and have larger uncertainty bands.
Our final results for the uncorrelated assumption (figure \ref{fig:Emax_3P1_k_3_compare_kmax_fixed_crms_0p6}(a)) are $C_{3P1}=2.8\pm0.0$, $D_{3P1}=0.5\pm0.2$ 
which is consistent with EKM's $\Emax=200$ MeV numbers $C_{3P1}=2.75$,
$D_{3P1}=0.32$, but is obtained without needing to specify a maximum energy for the fit~\footnote{Here we quote the marginalized one-dimensional 68\% intervals for the individual LECs. As discussed in section~\ref{sec:posteriors} the output of our analysis is, in fact, the full two-dimensional distribution for these LECs, which contains more information than this.}. The corresponding result for the correlated
assumptions (figure \ref{fig:Emax_3P1_k_3_compare_kmax_fixed_crms_0p6}(b)) is
$C_{3P1}=2.7\pm0.1$, $D_{3P1}=0.3\pm0.3$, which is statistically consistent
with the results in figure~\ref{fig:Emax_3P1_k_3_compare_kmax_fixed_crms_0p6}(a).

\begin{figure}[tbh]
  \centering
  \includegraphics[width=0.5\textwidth]{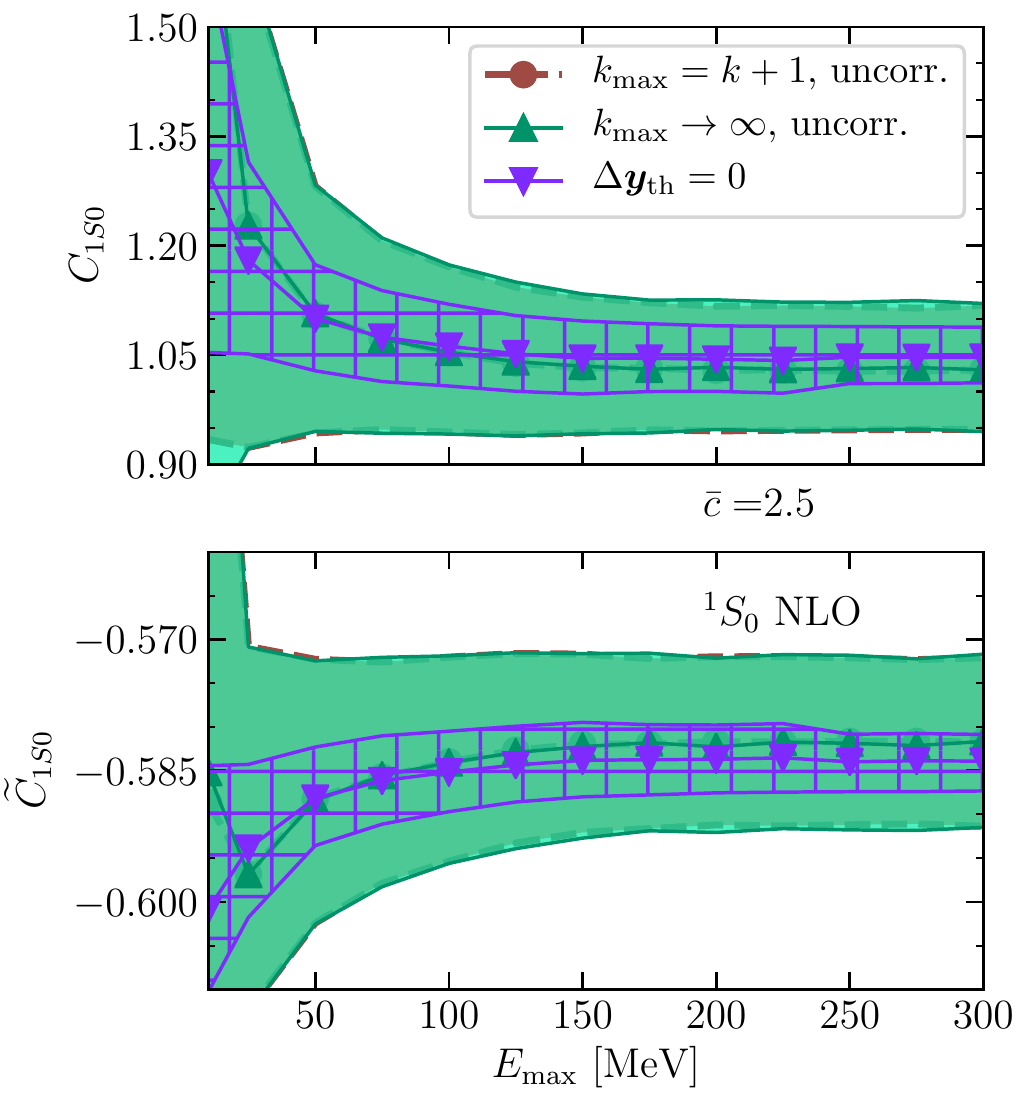}%
  \includegraphics[width=0.5\textwidth]{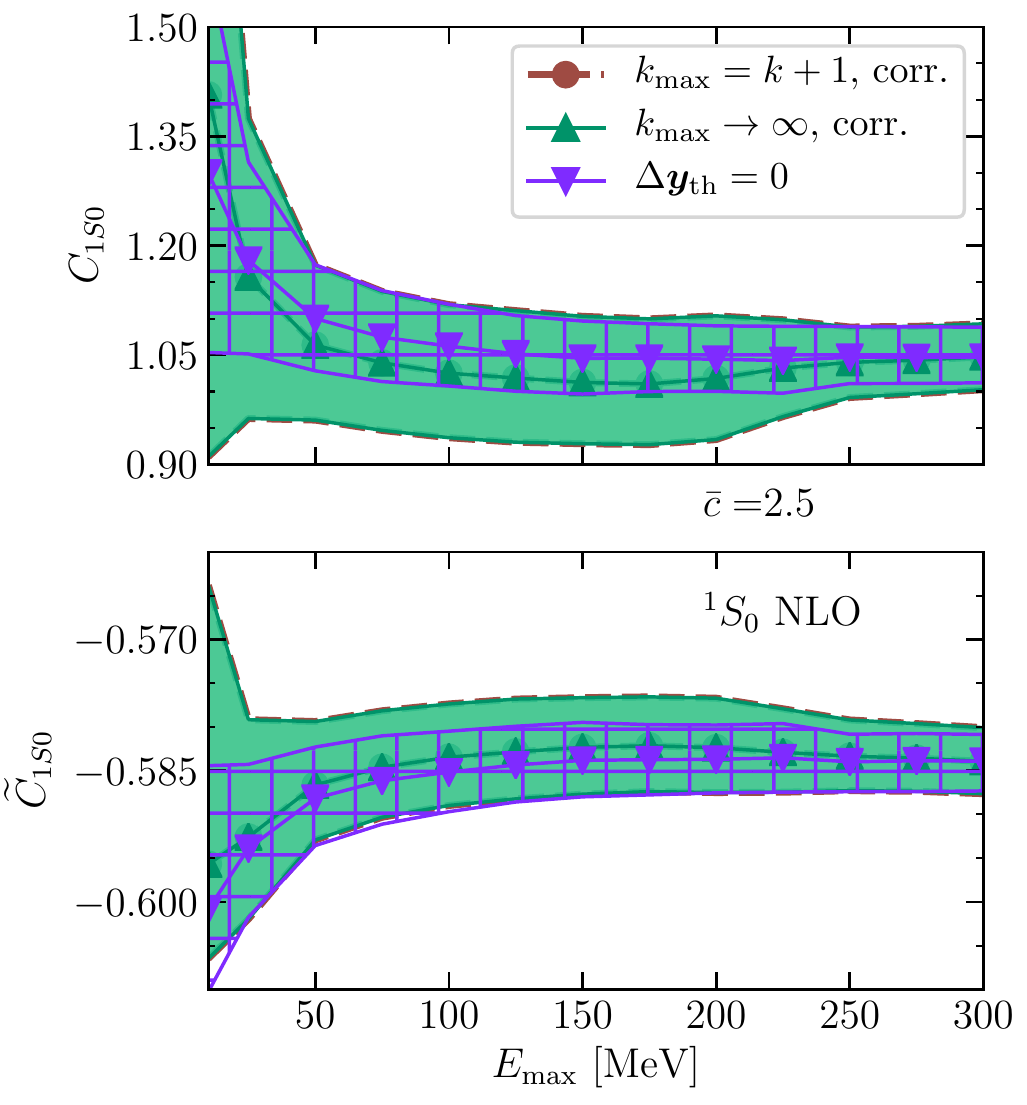}%
  \caption{$\Emax$ plots for the \NLO\ fits in the \oneSzero\  channel to
  the partial wave cross section.
  The no-truncation-error results ($\Delta\genobsvecth = 0$, purple inverted triangles) are compared 
  with those adding the theoretical error using the (a) uncorrelated and (b) fully correlated assumptions as in~\eqref{eq:covariance_case_A} and \eqref{eq:covariance_case_B}, using both the first-omitted-term
  approximation (coral circles) and the full $\kmax\to\infty$ 
  result (green triangles). The value of $\cbar = \sigmaemp=2.5$.
  \label{fig:Emax_1S0_k_1_compare_kmax_fixed_crms_2p46}}
\end{figure}

Finally, we look at another two-parameter case, the \oneSzero\ channel at \NLO, in figure~\ref{fig:Emax_1S0_k_1_compare_kmax_fixed_crms_2p46}.
Here the least-squares result looks stable for a large range of $\Emax$, but the implication of the
truncation error results is that we are seeing underfitting masquerading as a precise and accurate result.
We also note that the bands including truncation errors, while still consistent in the two limits, now
are noticeably larger in the uncorrelated case.

In summary, this case study highlights the critical importance of including EFT truncation errors to avoid
underfitting and achieve stable results for LECs, with error bands that reflect theoretical uncertainties
inherent in an EFT expansion.
While the size of error bands are exaggerated in this particular application by the prescription for
data errors of partial wave cross sections, the general lesson is robust: a least-squares fit with
only experimental errors will markedly underestimate the LEC uncertainty. 
This is particularly clear for the uncorrelated limit of the truncation error, where the theory error adds in quadrature to the experimental error.
The asymptotic stability with respect to the range of energies used in the fit (i.e.\ choice of $\Emax$) is a natural 
consequence of the energy-dependent weighting of the variances for the theory error in 
\eqref{eq:covariance_case_A} and \eqref{eq:covariance_case_B}; this discounts the influence of the
less-reliable higher-energy EFT calculations.
In the examples considered here, there were differences between the posteriors obtained under two different
limiting assumptions for the correlation structure of the theory errors with energy. But 
the LEC values obtained under both assumptions were always statistically consistent.
The actual dependence of the expansion coefficients 
$c_n$ on $\Elab$ clearly lies between these limits, strongly motivating a treatment 
with Gaussian processes that can take into account the finite correlation length in $\Elab$~\cite{GPpaper:2018}.

\section{Summary and Outlook} \label{sec:summary}

In this paper we explore the fitting of EKM's semi-local \eft\ \NN\ interactions~\cite{Epelbaum:2014efa,Epelbaum:2014sza} 
to 
selected \NN\ partial-wave phase shifts~\footnote{We recognize that the PWA93 phase shifts we use here as ``the data D" are not actually data, but are 
instead extracted from data under general assumptions. Our ultimate goal is to apply this framework to 
cases where $D$ is truly experimental, see also the discussion of future work below.}.
Our use of the term ``fitting'' belies the fact that we actually sample posteriors for the LECs in the \NN\ potential 
rather than simply maximising likelihoods. 
The posteriors are computed via simple formulae that follow from different simplifying assumptions. 
We present two case studies within our Bayesian parameter estimation framework~\cite{Wesolowski:2015fqa} that illuminate  features of parameter estimation which may be obscured or completely missed by 
conventional fitting protocols. This demonstrates the value of a Bayesian methodology even in the \NN\ case, where there are many precise data up to high energies. 
The case studies can be understood independently from the statistical background and derivations,
so we have separated those into appendices so as not to distract the reader.

Case study 1 (section~\ref{sec:posteriors}) illustrates the usefulness of projected posterior plots
to visualize multidimensional LEC pdfs.  
Such plots are important for understanding the full information content of the data.
In most channels examined in this section, both the one-dimensional and two-dimensional projections suggest Gaussian distributions 
(see figures~\ref{fig:phaseshift_1P1_N2LO_Emax_100MeV}--\ref{fig:phaseshift_post_3S1_NLO_Emax_100MeV}),
but a statistical test for normal distributions should always be applied
before Gaussian approximations for error propagation are used. 
We note that the occurrence of strong correlations between LECs, such as in figure~\subref*{fig:phaseshift_post_1S0_NLO_Emax_100MeV},
does not mean the posterior is non-Gaussian but implies that rotating the operator basis 
(e.g.\ by diagonalizing the covariance matrix) may be advisable.
The use of projected posterior plots as a physics diagnostic is dramatically illustrated by the projected
posteriors for the fourth-order $s$-wave LECs (figure~\ref{fig:phaseshift_post_1S0_N3LO_Emax_200MeV}), which
show strong deviations from normality 
despite providing a good fit to phase shifts. 
We show that this is a consequence of a parameter degeneracy at this order (see also \cite{Reinert:2017usi}).
This degeneracy was previously recognized in other contexts \cite{Furnstahl:2000we}, but was originally
discovered for \eft\ fits by looking at projected posteriors \cite{Wesolowski:2016int}.

Case study 2 (section~\ref{sec:Emaxplots}) addresses the question of the stability of LEC estimation
against the maximum energy of data used in the fit.  In non-Bayesian EFT parameter fitting, this 
choice of energy range can be a challenge: what is the optimal trade-off between including more data to
determine LECs more precisely and fit contamination from the increasing contributions at higher energies
of omitted higher-order EFT terms?  This case study shows that the
sensitivity to the choice of \Emax\ is removed with proper Bayesian uncertainty quantification.
The key is to account for truncation errors in the LEC posterior pdfs by adopting a suitable model discrepancy function.
This stops the LECs being unduly influenced by data at energies where the EFT order is too low to provide an accurate description.

The panels in figures~\ref{fig:Emax_1P1_k_1_compare_kmax_fixed_crms_0p89}, \ref{fig:Emax_1P1_k_2_compare_kmax_fixed_crms_0p89}, \ref{fig:Emax_3P1_k_3_compare_kmax_fixed_crms_0p6} and
\ref{fig:Emax_1S0_k_1_compare_kmax_fixed_crms_2p46} illustrate the desired behaviour for several partial 
waves at different orders for two limiting models of the truncation error: uncorrelated between different energies and 
the same for all energies (``fully correlated"). 
The approach to stability in \Emax\ varies in these examples, but in all cases we emphasize the spurious 
precision deduced when truncation errors are not accounted for (compare the $\Delta\genobsvecth = 0$  and $\kmax\rightarrow\infty$
curves in each figure). 
Note that the sensitivity to \Emax\ for $\Delta\genobsvecth = 0$  can be missed
because the quality of the fit as measured by a conventional least-squares metric may not be equally 
sensitive.
However, for a well-formulated EFT
the LECs are coefficients of an operator expansion, and therefore LEC values---although they may depend on the renormalization scheme
and scale used to define that expansion---should not, within errors, differ when different data are used to infer them.

While these case studies are for special \NN-only examples, the lessons are more general and will be
relevant for parameter estimation in the \NNN\ and \piN\ sectors,
where the quality and quantity of data is far less than in the \NN\ sector.
We also emphasize that, even with the restriction to the \NN\ sector, the task of fully applying 
Bayesian methods to \eft\ has many facets, which are quite entangled with one another.
We have only considered selected aspects here. 
Some other aspects that we are actively investigating---in concert with other groups---are:
\begin{itemize}
   \I generalizing the limiting models \eqref{eq:covariance_case_A} and \eqref{eq:covariance_case_B}
   used here for the correlation structure via a Gaussian process model~\cite{GPpaper:2018};
  \I using \NN\ scattering observables instead
  of extracted partial-wave phase shifts for parameter estimation;
  \I propagating all sources of error, including LEC uncertainties,
     to few- and many-body observables \cite{Furnstahl:2014xsa};
  \I accounting for correlations between LECs from the $\pi$N, 
     \NN\ and few-body sectors \cite{Carlsson:2015vda};
  \I estimating the \eft\ expansion parameter from the expected convergence
     pattern of observable predictions (this was explored for \NN\ in
     \cite{Furnstahl:2015rha,Melendez:2017phj});
  \I studying how to best parametrize the crossover between the $p/\Lambda_b$ and $m_\pi/\Lambda_b$
      expansions;    
  \I identifying appropriate priors for incorporating 
   other theoretical expectations such as Wigner symmetry;
 \I using Bayesian model selection to assess the impact of available experimental data on the number
     of orders of the EFT that can be constrained~\cite{Wesolowski:2015fqa};
  \I employing Bayesian model checking techniques to verify that the EFT expansion is working ``as advertised''. 
\end{itemize}

\section*{Acknowledgements}

We thank E.~Epelbaum, N.~Klco, S.~K\"onig, and 
R.~Navarro~P\'erez for useful discussions. 
The work of SW, RJF, and JAM was supported in part by the US National Science Foundation under
Grant No. PHY-1614460 and the NUCLEI SciDAC Collaboration under
US Department of Energy MSU subcontract RC107839-OSU. The work of SW was also supported
in part by Salisbury University.
The work of JAM was also supported in part by the U.S. Department of Energy, Office of Science, Office of Workforce Development for Teachers and Scientists, Office of Science Graduate Student Research (SCGSR) program. The SCGSR program is administered by the Oak Ridge Institute for Science and Education for the DOE under contract number DE‐SC0014664.
The work of DRP was supported by the US Department of Energy under
contract DE-FG02-93ER-40756 and by the 
 ExtreMe Matter Institute EMMI
at the GSI Helmholtzzentrum f\"ur Schwerionenphysik, Darmstadt, Germany.

\appendix

\section{Details for chosen EFT interaction} \label{sec:EFT-setup}

In this appendix we summarize the details of the EKM chiral interaction relevant to
the parameter estimation  of \NN\ LECs carried out in section~\ref{sec:posteriors}. 
Modern potentials derived from \eft\ employ different regulators and different parameter-estimation strategies~\cite{Carlsson:2015vda,Epelbaum:2014efa,Epelbaum:2014sza,Piarulli:2014bda,Reinert:2017usi,Entem:2017gor}.
It is not our purpose here to tackle issues for all of these. 
Instead, in section~\ref{sec:posteriors} 
we focused on estimating LECs from data, and concentrated on  the EKM semi-local coordinate-space chiral interaction up to 
\NNNNLO~\cite{Epelbaum:2014efa,Epelbaum:2014sza}. 

We examined the neutron-proton version of this interaction for simplicity. 
This interaction has local long-range
pion parts and non-local Gaussian contact interactions. If the expansion
is correctly renormalized, we expect a systematic convergence pattern
for observable calculations. In general, the expected convergence pattern
for an \NN\ observable calculated to order $n$ in chiral EFT is:
\begin{equation}
\genobs_n(p;\LEC_n)=\Xref \sum_{j=0}^n c_j Q^j; \quad n=0,\ldots, \kord;
\label{eq:csandts2}
\end{equation}
where $\genobs_{n}(p;\LEC_{n})$ is the $n$th order \eft\ prediction for the observable, $p$ is the momentum of interest, and we make explicit the
dependence of $\genobs_{n}$ on the set of LECs that appears at $n$th order, $\LEC_{n}$. 
Once we are in possession of 
these results for $\genobs_{n}$ in the EFT at orders $0$, $1$, \ldots, $\kord$, we can 
reconstruct the dimensionless coefficients $c_j$ that
define $\genobs$'s EFT expansion as in \eqref{eq:coefficient_definition}. 
This also requires that we identify the reference scale $\Xref$ and 
 the EFT expansion parameter, which here is $Q=\max\{p,m_\pi\}/\Lambda_b$, with 
$\Lambda_b$ the breakdown scale of the expansion. 
This particular specification of $Q$ is postulated, not derived; 
alternative prescriptions for the crossover between the $p/\Lambda_b$ and $m_\pi/\Lambda_b$ 
expansions are under investigation.
We note that in \eft\ the combination 
of parity and chiral symmetries guarantees that the coefficient $c_1$ is zero for all \NN\ scattering
observables. Therefore, while we report $c_1$ as part of the expansion (\ref{eq:csandts2}), $c_1$ provides
no information on the natural size of non-zero coefficients in the EFT expansion.
(Note: elsewhere we suppress $p$ and
use $\genobsvecn$, $\genobsrefvec$, and $\Qvec$ to denote sets of these
quantities at specified kinematic points.)

EKM follow Weinberg's power-counting prescription to organize contributions
to the \NN\ potential~\cite{Weinberg:1990rz,Weinberg:1991um}. In such an approach the expansion (\ref{eq:csandts2}) will only
hold for $\genobs_{n}$ if regulator artefacts are small. Therefore, the regulators in modern \eft\ potentials are chosen to minimize artefacts, 
which is important because our analysis
depends on the expected convergence. We explored this issue in the context of \eft\ truncation errors in \cite{Melendez:2017phj,Furnstahl:2015rha}, 
and demonstrated that---for the EKM semi-local interactions under discussion in this work---the regulator choice $R=0.9\,\mathrm{fm}$ produces
\NN\ observables that converge according to  (\ref{eq:csandts2}). However, this demonstration was only  
for observables summed 
over all partial waves: the convergence pattern (\ref{eq:csandts2}) is somewhat less systematic in individual partial 
waves. 
EKM determined the approximate breakdown
of the expansion using this cutoff to be $\Lambda_b = 600\,\mathrm{MeV}$, and we validated this 
result in  \cite{Furnstahl:2015rha,Melendez:2017phj} using Bayesian model checking. $\Lambda_b=600$ MeV is 
thus the value we take when computing $Q$ in  (\ref{eq:csandts2}). 

The value of $\Lambda_b$ also informs the expectation we have that the 
LECs which appear in the \NN\ potential will be of natural size. According to naive dimensional
analysis (NDA), we
expect the LECs that multiply \NN\ contact interactions will scale as~\cite{Epelbaum:2014efa,Epelbaum:2014sza}
  \beq
    |\widetilde{C}_i| \sim \frac{4\pi}{F_\pi^2}      \;,
    \qquad
    |C_i| \sim \frac{4\pi}{F_\pi^2 \Lambda_b^2}  \;,
    \qquad
    |D_i| \sim \frac{4\pi}{F_\pi^2 \Lambda_b^4}  \;,
    \label{eq:naturalness}
  \eeq
where the $\tilde C_i$s are \NN\ contacts at LO, the $C_i$s at NLO, and
the $D_i$s at \NNNLO. EKM report their $\tilde C_i$'s,  $C_i$'s and  $D_i$'s in
$10^4\,\text{GeV}^{-2}$, $10^4\,\text{GeV}^{-4}$, and $10^4\,\text{GeV}^{-6}$ respectively.
We work with explicitly dimensionless LECs in the \NN\ potential, which we obtain by removing from the EKM LECs
the dimensionless factors in  \eqref{eq:naturalness}. 
The LECs that appear in the different partial waves---and the associated partial-wave-basis momentum-space operators---are listed in table~\ref{tab:labels-lecs}. 

\begin{table}\centering
    \renewcommand{\arraystretch}{1.25}
    \caption{\label{tab:labels-lecs} 
    The partial-wave matrix elements of contact operators, up to N$^3$LO in the \eft\ expansion. 
    The two $\tilde C_i$'s are the LO ($k=0$) LECs. These are augmented by seven $C_i$'s in
    NLO fits, to which are added a further 15 $D_i$'s at \NNNLO\ (but cf.\ section \ref{sec:swave-case}).}
    \vspace{1mm}
    \begin{tabular}{@{}cc@{}}
    \br
     Partial wave  &  Scaled potential matrix element \\
     \mr
     \oneSzero\  & $\widetilde{C}^{np}_{1S0} + C_{1S0}(p^2 + \ppsq) + D^1_{1S0}\,p^2\,\ppsq + D^2_{1S0}(p^4 + \ppfo)$ \\
     \threeSone\ & $\widetilde{C}_{3S1} + C_{3S1}(p^2 + \ppsq) + D^1_{3S1}\,p^2\,\ppsq + D^2_{3S1}(p^4 + \ppfo)$ \\
     \onePone\  & $C_{1P1}\,p\,\pp + D_{1P1}\,p\,\pp\,(p^2 + \ppsq)$ \\
     ${}^3P_1$ & $C_{3P1}\,p\,\pp + D_{3P1}\,p\,\pp\,(p^2 + \ppsq)$ \\
     \threePzero\  & $C_{3P0}\,p\,\pp + D_{3P0}\,p\,\pp\,(p^2 + \ppsq)$ \\
     ${}^3P_2$ & $C_{3P2}\,p\,\pp + D_{3P2}\,p\,\pp\,(p^2 + \ppsq)$ \\
     ${}^1D_2$ & $D_{1D2}\,p^2\,\ppsq$ \\
     ${}^3D_2$ & $D_{3D2}\,p^2\,\ppsq$ \\
     ${}^3D_1$ & $D_{3D1}\,p^2\,\ppsq$ \\
     ${}^3D_3$ & $D_{3D3}\,p^2\,\ppsq$ \\
     ${}^3S_1-{}^3D_1$ & $C_{3S1-3D1}\,p^2 + D^1_{3S1-3D1}\,p^2\,\ppsq + D^2_{3S1-3D1}\,p^4$ \\
     ${}^3D_1-{}^3S_1$ & $C_{3S1-3D1}\,\ppsq + D^1_{3S1-3D1}\,p^2\,\ppsq + D^2_{3S1-3D1}\,\ppfo$ \\
     ${}^3P_2-{}^3F_2$ & $D_{3P2-3F2}\,p\,\ppcu $ \\
     ${}^3F_2-{}^3P_2$ & $D_{3P2-3F2}\,\pp\,p^3 $ \\
     \br
    \end{tabular}
\end{table}

Following EKM, we 
take the \piN\ coupling constants ($c_i$s) as given from a (particular) \piN\ 
analysis,\footnote{
  Note that the \piN\ coupling constants, conventionally defined as
   the $c_i$s, $d_i$s, etc., are not the same as the 
  observable coefficients ($c_n$s) we defined in  \eqref{eq:csandts2}.
} 
without propagating any uncertainties from that fit. Specifically, we
use the central value of the \piN\  couplings determined in~\cite{Epelbaum:2014sza}. In a future more complete 
analysis we should include these uncertainties in addition to those of the \NN\
LECs. Note that taking the \piN\  LECs as fixed in the \NN\ fit is in contrast to the simultaneous 
fit  to \piN\  and \NN\ data done by Carlsson \emph{et al}.~\cite{Carlsson:2015vda}.

\section{Bayesian methods} \label{app:bayes}

\subsection{Overview} \label{app:bayes_overview}

There are three main sources of uncertainty in EFT calculations, coming from
\bI
  \I truncating the EFT Lagrangian at a certain order, inducing systematic
  truncation errors \cite{Furnstahl:2015rha};
  \I fitting the free parameters of the EFT, the LECs, to data with experimental
  uncertainties;
  \I calculating the observables using a numerical method, e.g.\ a many-body
  calculation using an input nuclear potential.
\eI
These were outlined and discussed in detail in \cite{Furnstahl:2014xsa}. 
This is not an exhaustive list of uncertainties in EFT calculations. \eft\ 
in particular also has been demonstrated to result in predictions that
contain regulator artefacts~\cite{Dyhdalo:2016ygz}. The various sources of uncertainty
can also become entangled.

Other issues that arise in EFT parameter estimation deal with the quality of
how the fit describes the data. Particular issues are
\bI
  \I Overfitting. The model is finely tuned to the data. The goodness-of-fit
  increases with the model complexity, but this is because the model is tuned to data fluctuations. 
  This can occur in EFT if there is a parameter redundancy at a given order, or if effects at different
  orders are tuned to cancel with one another. 
  \I Underfitting. The model is too simplistic to describe the data. This can
  occur when an EFT truncated at low order is fit
  to data, some of which corresponds to a large value of the expansion parameter.
\eI
These issues are explored in a range of model problems in the context of
Bayesian model selection in \cite{Wesolowski:2015fqa}.

\subsection{Posterior pdf for LECs}\label{subsec:appendix_posterior_pdf}

The main goal of the parameter estimation is the calculation
of a joint posterior pdf for the LECs given a set of experimental data and any other information we have. 
The goal now is to use the rules of probability theory to
       express this posterior pdf in terms of known prior information so  we can calculate it. 
In a Bayesian framework, this task begins with a statistical model, as presented in \eqref{eq:stat_model}.
We can decompose that model further by introducing the true result $\genobsvec_{\mathrm{true}}$:
\begin{align}
    \label{eq:stat_model_alt}
    \genobsvec_{\mathrm{true}} =  \genobsvecthk + \Delta\genobsvecthk \;;
    \qquad
    \genobsvecexp = \genobsvec_{\mathrm{true}} + \Delta\genobsvecexp \;.
\end{align}
Here $\genobsvecthk$ is a set of EFT results (\ref{eq:csandts2}) computed at a particular order $k$ 
for various momenta $p_i$: $\genobsvecthk =\{y_k(p_i;\LECk):i=1,\ldots,N_d\}$, with $N_d$  the 
total number of data points considered. $\genobsvecexp$ is then the corresponding vector of experimental 
results at these momenta, which comes together with a statistical model
for the experimental errors $\Delta \genobsvecexp$. The expansion parameter $Q$ also depends on the 
kinematic point, as does the reference scale for the observable. 
These are collected in vectors $\Qvec$ and $\genobsrefvec$. 

\begin{figure}[tbh]
  \centering
  \includegraphics[width=0.8\textwidth]{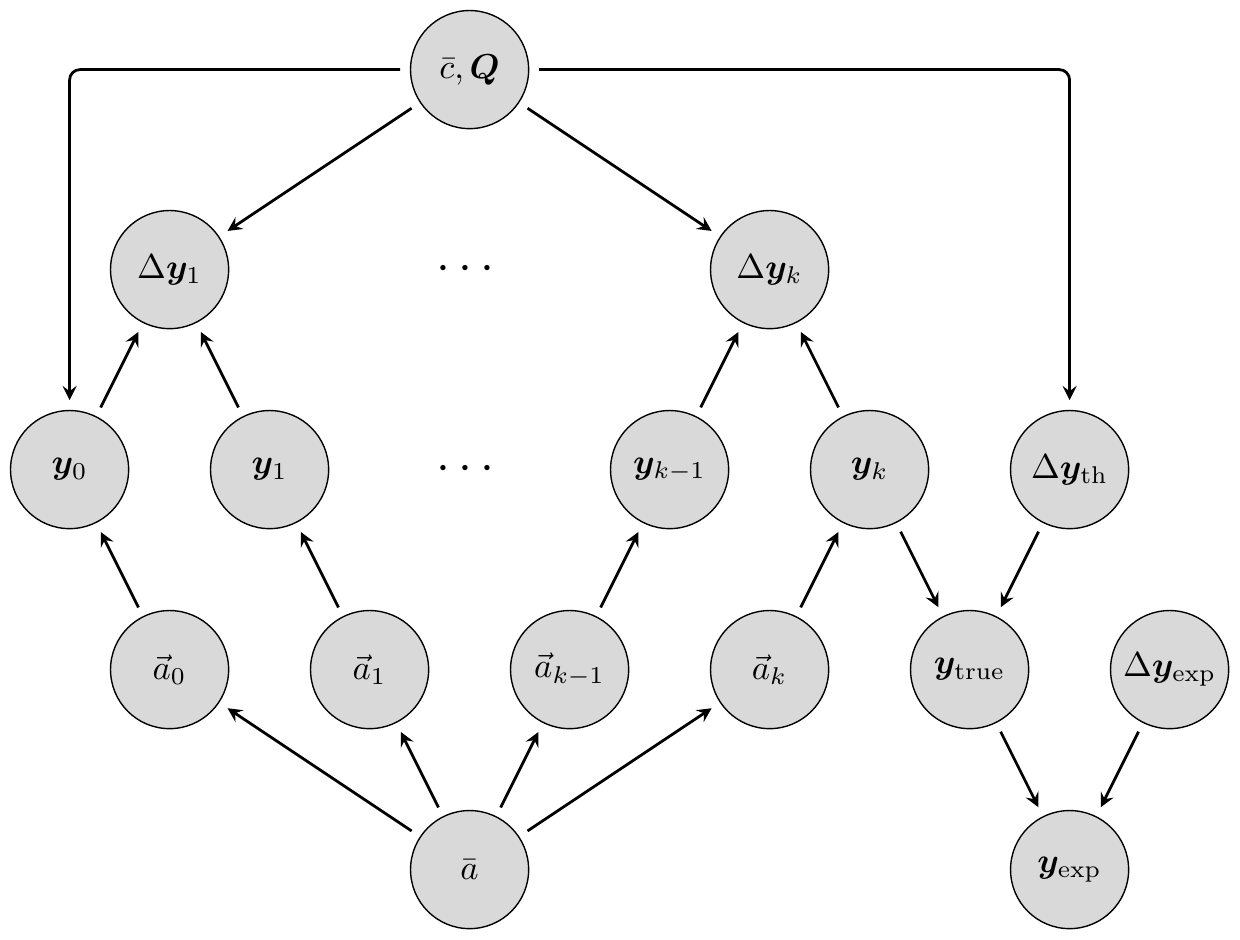}
  \caption{Bayesian network for the model \eqref{eq:stat_model_alt}. The differences between orders $\Delta\genobs_n$ are defined in~\eqref{eq:coefficient_definition}.
  The top node implicitly contains $\genobsrefvec$ and the assumption of correlation structure of the $\cvec_n$.}
  \label{fig:Bayesian_network}
\end{figure}

In this paper we employ two limiting models for discrepancy $\Delta \genobsvecthk$, as described in Sec.~\ref{subsec:posterior_pdf}.
A Bayesian network~\cite{ben2007bayesian,citeulike:13925803} 
for \eqref{eq:stat_model_alt} is given in figure~\ref{fig:Bayesian_network}.
We will show how writing our model as~\eqref{eq:stat_model_alt} and explicitly writing a Bayesian network as in figure~\ref{fig:Bayesian_network} can help us derive the posterior for $\LECk$ in two different ways.

For estimating LECs in \eft, we write the posterior pdf\footnote{%
  The quantity $\pr(x \given y,I)$ is read ``the probability that $x$ is true given
  $y$ and other background information $I$.''
} of interest as
\begin{align}
  \pr(\LECk \given \genobsvecexp, I) = 
    \frac{\pr(\genobsvecexp \given \LECk, I) \pr(\LECk\given I)}{\pr( \genobsvecexp \given I )} \;
  \;,
\end{align}
where we have used Bayes theorem\footnote{This work does not contain a comprehensive introduction
to Bayesian methods, but builds on the formalism developed in \cite{Wesolowski:2015fqa}.
For textbook introductions to Bayesian parameter estimation and other applications,
we refer the interested reader to \cite{Sivia:2006,gelman2013bayesian,Gregory:2005}.} 
to write it in terms of the \emph{likelihood} for the data given the LECs, 
$\pr(\genobsvecexp \given \LECk, I)$,
times the \emph{prior}, $\pr(\LECk\given I)$, which is the probability distribution
of the LECs given only the background information that we have, that is, not
mediated by any measured data.
The term in the denominator, $\pr( \genobsvecexp \given I )$, is known as the \emph{evidence}
or the \emph{marginal likelihood}. It does not explicitly depend on the LECs $\LECk$
and is just a normalization factor for our purposes. 
When the data constrains the likelihood well, the form of the prior 
has little to no impact on the final results.

 Additional information $I$, such as the variance
of the experimental data, will be specified explicitly as needed.
Our prior for $\LECk$ is based on the idea that if the relevant
physical scales are identified, the dimensionless LECs should all be about the same magnitude
but statistically independent of each other.
We incorporate naturalness via a Gaussian prior on $\LECk$, whose standard deviation $\abar$ is of order unity
\begin{align}
  \pr(\LECk \given \abar) = \biggl(\frac{1}{\sqrt{2\pi}\abar}\biggr)^{\nlecs} \eup^{-[\LECk]^2/2\abar^2}
  \;.
\end{align}
Here, $\nlecs$ is the number of LECs in $\LECk$ and $[\LECk]^2$ is the sum of the squares of
the LECs.
In this work we will use a fixed value of $\abar$. 
More generally, the prior for $\LECk$ unconditional on $\abar$
would be obtained by proposing another prior $\pr(\abar)$ and marginalizing:
\begin{align}  \label{eq:abar_marginalization}
  \pr(\LECk) = \int_0^{\infty} \dd{\abar} \pr(\LECk\given \abar) \pr(\abar)
  \;.
\end{align}
However, the role of the prior on LECs is simple: impose a penalty on any LEC that is too large, which (generally)
indicates overfitting.
Previous work has shown little sensitivity to the functional form of this prior and to the value of $\abar$
within a reasonable range (see \cite{Furnstahl:2014xsa} and \cite{Wesolowski:2015fqa} for model examples).

\subsection{Interlude: fun with Gaussian integrals, delta functions and matrices}

To complete the derivation of $\pr(\LECk \given \genobsvecexp, I)$, it is useful to remember the following relations.
First, the Dirac $\delta$ function can be written as a Fourier integral
\begin{align} \label{eq:fourier_delta}
  \delta(x - y) = \intfull \frac{\dd{\omega}}{2\pi} \, e^{i\omega(x-y)}  \;,
\end{align}
which is convenient when $x$ and $y$ are intermediate dummy variables in a larger integral.
Additionally,~\eqref{eq:fourier_delta} will be used to prove the second relation---the well-known sum rule for Gaussian random variables.
Let's consider two independent Gaussian random variables, $X$ and $Y$, which we notate as
\begin{align}
  X & \sim \normal(\mu_x, \sigma_x^2) \;, \\
  Y & \sim \normal(\mu_y, \sigma_y^2) \;,
\end{align}
where the notation $\normal(\mu,\sigma^2)$ will be used extensively in what follows to denote a normally distributed random variable with mean $\mu$ and standard deviation $\sigma$, i.e.
\begin{align}  \label{eq:Gaussian_pdf}
  X \sim \normal(\mu_x,\sigma_x^2) 
    \quad \Longleftrightarrow \quad
  \pr(x|\mu_x,\sigma_x) = \frac{1}{\sqrt{2\pi}\sigma_x} \eup^{-(x-\mu_x)^2/2\sigma_x^2}
  \;.
\end{align}
The dependence on the mean and standard deviation $X \given \mu_x, \sigma_x$ 
is frequently omitted for the sake of brevity. 
The sum of $X$ and $Y$ is then distributed as
\begin{align} \label{eq:1d_gaussian_sum_rule}
  X + Y \sim \normal(\mu_x + \mu_y, \sigma^2_x + \sigma^2_y) \;.
\end{align}
Although~\eqref{eq:1d_gaussian_sum_rule} is well known, we derive it here as a primer for subsequent derivations. 

Define $Z$ as the sum of $X$ and $Y$
\begin{align}  \label{eq:Gaussian_sum}
  Z = X + Y \;.
\end{align}
We seek the pdf for $Z$ given the means and variances of $X$ and $Y$, which can be obtained by the following steps~\cite{Sivia:2006}
\begin{align}
  \pr(z) &= \intfull \dd{x} \dd{y} \pr(z,x,y) \tag{marginalization} \\
     &= \intfull \dd{x} \dd{y} \pr(z\given x,y) \pr(x,y) \tag{product rule} \\
     &= \intfull \dd{x} \dd{y} \pr(z\given x,y) \pr(x) \pr(y) \tag{independence} \\
     &= \intfull \dd{x} \dd{y} \pr(z\given x,y)
          \frac{1}{\sqrt{2\pi}\sigma_x} \eup^{-(x-\mu_x)^2/2\sigma_x^2}
          \frac{1}{\sqrt{2\pi}\sigma_y} \eup^{-(y-\mu_y)^2/2\sigma_y^2}
        \;. \label{eq:first_integrate_in}
\end{align}
Note that, given $x$ and $y$, $z$ is exactly known via~\eqref{eq:Gaussian_sum}.
This distribution is expressed as a Dirac delta function and its corresponding Fourier integral
\begin{align} \label{eq:Fourier_delta_function}
  \pr(z\given x,y) = \delta(z - x - y)
      = \frac{1}{2\pi} \intfull \dd{\omega} \eup^{-i\omega z} \eup^{i\omega x} \eup^{i\omega y}
           \;.
\end{align}
Though~\eqref{eq:first_integrate_in} could be evaluated using the standard form of the $\delta$ function, the Fourier integral is readily generalizable to the other relations in this work.
Substituting \eqref{eq:Fourier_delta_function} into \eqref{eq:first_integrate_in}, we see that the
$x$ and $y$ integrals factor: 
\begin{align} \label{eq:factored_integrals}
  \pr(z) = \frac{1}{2\pi}\intfull \dd{\omega} \eup^{-i\omega z}
    &\Bigl[ \frac{1}{\sqrt{2\pi}\sigma_x} \intfull \dd{x} \eup^{i\omega x} \eup^{-(x-\mu_x)^2/2\sigma_x^2} \Bigr] 
      \notag \\
  \null\times  
    &\Bigl[ \frac{1}{\sqrt{2\pi}\sigma_y} \intfull \dd{y} \eup^{i\omega y} \eup^{-(y-\mu_y)^2/2\sigma_y^2} \Bigr] 
    \;.
\end{align}
The terms in brackets are just Fourier transforms of Gaussian distributions:
\begin{align} \label{eq:Fourier_transform_Gaussian1}
   \intfull \dd{t} \eup^{i\nu t} \eup^{-(t-\mu)^2/2\sigma^2} &=  
      \sqrt{2\pi}\sigma \eup^{i\nu\mu} \eup^{-\sigma^2\nu^2/2} \;, \\
   \frac{1}{2\pi}\intfull \dd{\nu} \eup^{-i\nu (t-\mu)} \eup^{-\sigma^2\nu^2/2} &=  
    \frac{1}{\sqrt{2\pi}\sigma} \eup^{-(t-\mu)^2/2\sigma^2}
    \label{eq:Fourier_transform_Gaussian2}
   \;.
\end{align}
With the above relations, the integrals in \eqref{eq:factored_integrals} simplify to
\begin{align}
    \pr(z) &= \frac{1}{2\pi}\intfull \dd{\omega} \eup^{-i\omega z}
       \eup^{i\omega\mu_x} \eup^{-\sigma_x^2\omega^2/2} \eup^{i\omega\mu_y} \eup^{-\sigma_y^2\omega^2/2}
       \notag  \\
       &= \frac{1}{2\pi}\intfull \dd{\omega} \eup^{-i\omega (z - (\mu_x+\mu_y))}
            \eup^{-(\sigma_x^2+\sigma_y^2)\omega^2/2}  \notag \\
       &= \frac{1}{\sqrt{2\pi(\sigma_x^2+\sigma_y^2)}} \eup^{-\left(z-(\mu_x+\mu_y)\right)^2/2(\sigma_x^2+\sigma_y^2)}
       \;,
\end{align}
from which~\eqref{eq:1d_gaussian_sum_rule} follows.

More generally, given scalars $a$ and $b$,
\begin{align} \label{eq:weighted_gauss_sum}
  aX + bY \sim \normal(a\mu_x + b\mu_y, a^2\sigma^2_x + b^2\sigma^2_y)\;,
\end{align}
and, for $N$-dimensional independent random variables $\mathbf{X}$ and $\mathbf{Y}$ and $M\times N$ 
dimensional matrices $A$ and $B$,
\begin{align} \label{eq:weighted_mv_gauss_sum}
  A\mathbf{X} + B\mathbf{Y} \sim \normal(A\boldsymbol{\mu}_x + B\boldsymbol{\mu}_y, A\Sigma_xA^T + B\Sigma_yB^T) \;.
\end{align}
Finally, we note the Woodbury matrix identity for an $N\times N$ matrix $A$, $N \times M$ matrix $U$, $M \times M$ 
matrix $C$, and $M\times N$ matrix $V$:
\begin{align} \label{eq:woodbury}
  (A + UCV)^{-1} = A^{-1} - A^{-1} U (C^{-1} + V A^{-1} U)^{-1} V A^{-1}
  \;.
\end{align}
With the Fourier representation~\eqref{eq:fourier_delta}, the Gaussian sum rule in its various forms, and the Woodbury identity,
we are now ready to derive the likelihood, and hence the full posterior, for EFT parameter estimation.

\subsection{The tale of two likelihoods}

Before computing the likelihood $\pr(\genobsvecexp\given \LECk, I)$, it is helpful to derive how the theoretical discrepancy $\Delta\genobsvecth$ is distributed.
Consider the scalar quantity
\begin{align}
  \Delta \genobsth = \genobsref\sum_{n=k+1}^\infty c_n Q^n \;,
\end{align}
where
\begin{align}
  c_n \given \cbar \sim \normal(0, \cbar^2) 
\end{align}
and $Q$ is known with $0 \leq Q < 1$.
Then by~\eqref{eq:weighted_gauss_sum}, it follows that 
\begin{align}
  \Delta \genobsth \given \cbar, Q \sim \normal\left(0, \genobsref^2\cbar^2\frac{Q^{2(k+1)}}{1-Q^2}\right),
\end{align}
as derived in \cite{KlcoThesis:2015}. Note that we we will now explicitly condition on $\cbar$ and $Q$.
From our Bayesian network for \eqref{eq:stat_model_alt} in figure~\ref{fig:Bayesian_network}, this common
value of $\cbar$ governs (and hence can be inferred by) all of the coefficients $c_n$ at all of the kinematic points.  
A true Bayesian, following the enlightened path, would then compute
\begin{align}
  \pr(\cbar \given \ckvecset) \propto \pr(\ckvecset \given \cbar) \pr(\cbar)\;.
\end{align}
However, it is well known that given a large number of expansion coefficients, the root-mean-square
$\sigmaemp$, or sample standard deviation, is an accurate estimator of $\cbar$, 
and we can simply use that value to good approximation. 
Explicitly,
\begin{align} \label{eq:crms_equation}
   \sigmaemp = \sqrt{\frac{1}{N_{\rm tot}} \sum_{\{c_n\}} c_n^2} \;,
\end{align}
where $\{c_n\}$ is a representative set of $N_{\rm tot}$ extracted coefficients 
at different energies.

The generalization of $c_n$ from scalar to vector (or functional) quantities requires that we make assumptions 
about how the $c_n$s at different kinematic points are correlated. I.e.\ if we want to compute theoretical uncertainties
at a range of energies we need to understand how the errors at different kinematic points are correlated with one another.
In such a circumstance $\cvec_n$ is a vector in the space of momentum values $p_i$ and so should be thought of as a multidimensional 
random variable. What correlation structure does it have? Here we consider two limiting possibilities 
for the correlations between the elements of $\cvec_n$, $R_{ij} \equiv \Corr[\cvec_{n,i}, \cvec_{n,j}] = \Cov[\cvec_{n,i}, \cvec_{n,j}]/\cbar^2$, namely:
\begin{align}
  \label{eq:uncorrelated_case}
  \corrmat_{ij,\mathrm{uncorr.}} & = \delta_{ij} \;, \\
  \label{eq:fully_correlated_case}
  \corrmat_{ij,\mathrm{corr.}}  & = 1  \;,
\end{align}
where $n$ denotes the order of the coefficient and $i$ and $j$ index the momentum at which it is to be evaluated. 
The fully correlated case \eqref{eq:fully_correlated_case} assumes that these coefficients in the $Q$-expansion for the observable under consideration do not themselves depend on momentum;
the values of $c_n$ at different kinematic points are then 100\% correlated with one another. 
In contrast, in the uncorrelated case \eqref{eq:uncorrelated_case} we take $c_n$ at one momentum to be completely independent of $c_n$ at a different momentum; the values of $c_n$ at 
different kinematic points are statistically independent of one another. 
We note that Gaussian processes can describe these two extreme cases and intermediate correlation structures as well.
They will be explored in future work~\cite{GPpaper:2018}.

To derive the theoretical covariance for the multivariate case, we first rewrite our truncation error model as
\begin{align}
  \Delta \genobsvecth = \genobsref\sum_{n=k+1}^\infty Q^n \cvec_n \;,
\end{align}
where we have now promoted $\genobsref$ and $Q$ to $\Nexp\times \Nexp$ diagonal matrices.
Now we can write
\begin{align}
  \cvec_n \given \cbar \sim \normal(0, \cbar^2 \corrmat)  \;,
\end{align}
where we remain agnostic to the correlation structure $\corrmat$.
Now the general multivariate sum~\eqref{eq:weighted_mv_gauss_sum} can be applied repeatedly to show
\begin{align}
  \Delta \genobsvecth \given \cbar, Q \sim \normal\left(0, \covarth\right)  \;,
\end{align}
where
\begin{align}
  [\covartharg{corr.}]_{ij} & = \cbar^2 {\genobsref}_{ii}{\genobsref}_{jj} 
                                 \frac{(Q_{ii} Q_{jj})^{k+1}}{1 - Q_{ii} Q_{jj}} \;, \\
  [\covartharg{uncorr.}]_{ij} & = [\covartharg{corr.}]_{ij} \delta_{ij}  \;.
\end{align}
When specificity is not required, we will encapsulate both cases by using $\covarth$ as the theory covariance matrix.

As suggested by~\eqref{eq:stat_model_alt}, the distribution of the experimental data given the LECs (i.e., the likelihood) is given by the sum of the theoretical prediction, the theoretical uncertainty, and the experimental uncertainty.
Given the LECs, the prediction is exactly known, while we have shown that the theoretical uncertainty is a multivariate Gaussian.
Assuming that the experimental noise is also a Gaussian
\begin{align}    \label{eq:prior_for_data}
  \Delta \genobsvecexp \sim \normal(0, \covarexp) \;,
\end{align}
then it follows that the likelihood is
\begin{align} \label{eq:mv_likelihood_quadrature}
  \genobsvecexp \given \LECk, I \sim \normal(\genobsvecth, \covarth + \covarexp)  \;,
\end{align}
where $I = \cbar, \Qvec, \genobsrefvec, \covarexp$.
Hence, the covariances are additive.

In fact, after some algebra,~\eqref{eq:mv_likelihood_quadrature} could have been derived by integrating in 
the unknown value for $\Delta\genobsvecth$ at each input point
\begin{align}
  \pr(\genobsvecexp \given \LECk, I) = \intfull \pr(\genobsvecexp \given \LECk, \Delta\genobsvecth, Q, \covarexp) \pr(\Delta\genobsvecth \given \cbar, \Qvec) \dd[\Nexp]{(\Delta\genobsvecth)}  \;.
\end{align}
For the special case of the fully correlated coefficients and a finite $\kmax$, we could instead integrate in each (univariate) unknown coefficient~\cite{Stump:2001gu,Schindler:2008fh}:
\begin{align} \label{eq:likelihood_marg_all_cn}
  \pr(\genobsvecexp \given \LECk, I) = \intfull \pr(\genobsvecexp \given \LECk, \Delta\genobsvecth, \Qvec, \covarexp) \prod_{n=k+1}^{\kmax}\pr(c_n \given \cbar) \dd{c_n}  \;.
\end{align}
Interestingly, by following~\eqref{eq:likelihood_marg_all_cn} and completing the square, 
we arrive at a modified least squares likelihood that on the surface is completely different from~\eqref{eq:mv_likelihood_quadrature}.
This alternative form can be derived and understood in relation to~\eqref{eq:mv_likelihood_quadrature} 
by isolating $\covarexp$ via the Woodbury matrix identity~\eqref{eq:woodbury}.
In particular, we note that the following decomposition holds
\begin{align}
  \covartharg{corr.} & = U U^T  \;, \\
  U_{in} & = \cbar {\genobsref}_i Q_i^{k+n}  \;,
\end{align}
where $1 \leq i \leq \Nexp$ and $1 \leq n \leq \kmax-k$.
Therefore,
\begin{align}
  (\covarexp + \covartharg{corr.})^{-1} & = (\covarexp + U U^T)^{-1} \notag\\
  & = \covarexp^{-1} - \covarexp^{-1} U (\identity + U^T \covarexp^{-1} U)^{-1} U^T \covarexp^{-1} \;.
\end{align}
If we make the definitions
\begin{align}
  A & = \identity + U^T \covarexp^{-1} U  \;, \\
  \myvec{b} & = U^T \covarexp^{-1} [\genobsvecexp - \genobsvecth] \;, \\
  \chi^2 & = [\genobsvecexp - \genobsvecth]^T \covarexp^{-1} [\genobsvecexp - \genobsvecth] \;,
\end{align}
then a modified least squares likelihood---which is also derived, but without the use of the Woodbury identity in \cite{Stump:2001gu} and \cite{Schindler:2008fh}---results
\begin{align} \label{eq:modified_chi2_likelihood}
  \pr(\genobsvecexp \given \LECk, I, \mathrm{corr.}) \propto e^{-\chi^2/2} e^{\myvec{b}^T A^{-1}\myvec{b}/2} \;.
\end{align}
The form of~\eqref{eq:modified_chi2_likelihood} is potentially advantageous when $\kmax - k$ is small and 
$\covarexp$ is diagonal, since the size of the matrix inversion for $A$ is only $(\kmax - k) \times (\kmax - k)$ 
rather than $\Nexp \times \Nexp$ as in~\eqref{eq:mv_likelihood_quadrature}.


\section{Short-range test of operator redundancy}  \label{app:redundantpionless}

In this appendix we explore the issue of operator redundancy in the 
$s$-wave potential at $O(Q^4)$ in  
a theory with only short-range ({s.r.}) interactions. Here the behaviour is clearer because
there are no long-range (one-pion exchange, two-pion exchange, etc.) pieces of the interaction.
We demonstrate that---at least 
for the range of LEC values where the $O(Q^4)$ piece
of the potential can be treated in perturbation theory---the combination of LECs multiplying the operator that disappears on-shell can be absorbed into
lower-order LECs. 

We consider a 
pure short-range  theory with momentum-space cutoff regularization. This ``toy model" has 
a $s$-wave contact potential between nucleons given by
\beq
  \VSR(p,p^\prime) = \CLO + \Ctwo (p^2 + p^{\prime2} ) + \Cfo (p^2 + p^{\prime2} )^2
  + \Cft (p^2 - p^{\prime2} )^2 {}\;,
  \label{eq:VSR}
\eeq
to which we will apply a separable cutoff regulator $g(p)g(p^\prime)$.
For example, we could take $g(p) = e^{-p^2/\Lambda_c^2}$ as in the EKM chiral
interactions in \cite{Epelbaum:2014efa}.
We will treat the first two terms of \eqref{eq:VSR} nonperturbatively to
simulate our \eft\ problem from section~\ref{sec:swave-case}. 
 When the leading [$O(Q^0)$] and quadratic [$O(Q^2)$] are resummed as in \cite{Phillips:1997xu},
the off-shell $T$-matrix is given by
\begin{align}
    \Tzero(p,\pp; E=k^2/2\mu) 
    & = \frac{g(p)\, g(\pp)}{(\Ctwo I_2 - 1)^2 - \CLO I_0 - \Ctwo^2 I_0 I_4} \notag \\
    & \quad \null \times \left[ \CLO + \Ctwo^2 I_4 + (\Ctwo - \Ctwo^2 I_2)(p^2 + p^{\prime 2})
    +  \Ctwo^2 I_0 p^2 p^{\prime 2} \right] \;,
    \label{eq:Tmat-soln-leading}
\end{align}
where $\mu=M/2$ is the reduced mass in the \NN\ system. 
The regulated loop integrals $I_0$, $I_2$, and $I_4$ are functions of the cutoff
$\Lambda_c$ and the on-shell momentum $k$, and are defined by:
\begin{align}
  I_{2n}(\Lambda_c, k) \equiv \frac{4\mu}{\pi} \int_{0}^\infty \dd{q}
  \frac{q^{2+2n}\, g^2(q)}{k^2 - q^2 + i \epsilon} \;.
\end{align}
The on-shell T-matrix can be written:
\begin{align}
    \Tzero(k,k; E=k^2/2\mu) 
    & = \frac{g^2(k)}{(\Ctwo I_2 - 1)^2 - \CLO I_0 - \Ctwo^2 I_0 I_4} \notag \\
    & \quad \null \times \left[ \CLO + \Ctwo^2 J_5 + (2\Ctwo - \Ctwo^2 J_3)k^2 \right] \;,
    \label{eq:Tmat-leading-onshell}
\end{align}
where we have used $I_{2n} = k^2 I_{2n-2} + J_{2n+1}$, with $J_{2n+1}$ a purely divergent integral
that for cutoff regulators can be parameterized as:
\begin{align}
J_{2n+1} \equiv -\frac{4 \mu}{\pi} \frac{\beta_{2n+1}}{2n+1} \Lambda_c^{2n+1} \;.
\end{align}
The dimensionless numbers $\beta_{2n+1}$ should be of order one; indeed, they all equal to one if 
a sharp cutoff $g(p)=\theta(\Lambda_c - p)$ is used. 
This behaviour of the integrals then means that the dimensionful coefficients in the potential are
expected to scale with $\Lambda_c$ as
 $C_{2n} \sim 2 \pi/\mu \Lambda_c^{2n+1}$~\cite{Beane:1997pk,Kaplan:1998tg,Kaplan:1998we,vanKolck:1998bw}.

We now use perturbation theory to add the effects of the two $O(Q^4)$ terms to the $T$-matrix
so as to see the extent to which those operators affect the on-shell amplitude at this order. 
(See \cite{Gegelia:1998iu} for 
the $T$-matrix when the full potential \eqref{eq:VSR} is used in the Lippman-Schwinger equation.)
Note that this is not a standard pionless-EFT calculation, because there the $O(Q^2)$ term
would also be treated perturbatively. (It is also unusual to use cutoff regularization, because
the power counting is clearer when dimensional regularization with power-divergence subtraction
is employed~\cite{Kaplan:1998tg,Kaplan:1998we}.) If $\CLO$ and $\Ctwo$ are treated non-perturbatively
and $\Cft$ treated perturbatively, we obtain for the shift in the on-shell $T$-matrix: 
\begin{align}
  \Delta T^{C_{42}}(k,k;E) &= \frac{-2 \Cft g^2(k)}{[(\Ctwo I_2 - 1)^2 - \CLO I_0 - \Ctwo^2 I_0 I_4]^2}
               [ (\CLO^2 J_3^2 - \CLO J_5 + 3 \CLO \Ctwo J_3 J_5  \notag \\ 
       &\quad \null + \Ctwo^3 J_3 J_5^2- \Ctwo J_7 + 2 \Ctwo^2 J_3 J_7 - \Ctwo^3 J_3^2 J_7)  \notag \\ 
       &\quad \null + (\CLO J_3 + \CLO \Ctwo J_3^2  + 2 \Ctwo^2 J_3 J_5) k^2 + \Ctwo J_3 k^4] \;.
     \label{eq:pertfoshift1}
\end{align}

Only the power-law divergent integrals $J_3$, $J_5$, and $J_7$ appear in the numerator of (\ref{eq:pertfoshift1}):
the fact that $\Cft$ only affects the potential off-shell results in factors of $q^2 - k^2$  in all relevant 
integrals, which cancel the \NN\ propagator that is present in the integrals $I_{2n}$ defined above. 
The $J_{2n+1}$s are thus pure ultra-violet integrals that contain no infra-red physics and so cannot be predicted 
within the EFT: the low-energy propagation of \NN\ pairs is not present in them. Because of this, $J_{2n+1}=0$ for 
all $n \geq 0$ in dimensional regularization with minimal subtraction. Here we are using a cutoff, so they are not 
zero, but they should all be renormalized by LECs. This is feasible because the only terms that appear in the 
numerator of (\ref{eq:pertfoshift1}) are polynomial in $k^2$. 

We note two other things about  (\ref{eq:pertfoshift1}). First, because $\Cft$ is an off-shell effect it only 
affects $T(k,k;E)$ through terms that also involve the LECs $\CLO$ and $\Ctwo$. This is particularly notable for 
the $k^4$ term, which would not be present had we treated both $\Ctwo$ and $\Cft$ in perturbation theory. Second, 
one might be concerned about the denominator in  (\ref{eq:pertfoshift1}) being the square of that 
in  (\ref{eq:Tmat-soln-leading}). This, however, is a canonical consequence of our use of Distorted-Wave 
Born Approximation to 
evaluate the effects of $\Cft$. The ``$T^2$" piece of $(T G_0 + 1) V^{(4)} (1 + G_0 T)$ will generically produce 
such a structure. 

With this in mind, we proceed to renormalize the $k^0$ and $k^2$ terms in the numerator of  
(\ref{eq:pertfoshift1}) by including in the analysis a perturbative shift of both $C_0$ and $C_2$. 
We add a small correction to $\VSR$ that has exactly the same structure as the second-order potential:
\begin{equation}
   V^{({\rm LEC~shift})}(p,p')=\Delta \CLO + \Delta \Ctwo (p^2 + p^{\prime \, 2}) \;.
\end{equation}
Working to first-order in $\Delta \CLO$ and $\Delta \Ctwo$, we obtain for the resulting change in the on-shell $T$-matrix:
\begin{align}
 \Delta T^{\rm LEC~shift}(k,k;E) &= \frac{2\Delta \Ctwo g^2(k)}{[(\Ctwo I_2 - 1)^2 - \CLO I_0 - \Ctwo^2 I_0 I_4]^2}
      \notag \\
      & \quad \null \times [ (\CLO J_3 - \CLO \Ctwo J_3^2 + \Ctwo J_5 - \Ctwo^2 J_3 J_5) + (1 - \Ctwo J_3 )k^2] \notag \\
      & \quad \null +  \frac{\Delta \CLO g^2(k)}{[(\Ctwo I_2 - 1)^2 - \CLO I_0 - \Ctwo^2 I_0 I_4]^2}    
           [1 - 2 \Ctwo J_3 + \Ctwo^2 J_3^2]  \;.
       \label{eq:pertfoshift2}
\end{align}
Here we have terms proportional to $\Delta C_0$ and $\Delta C_2$ that can absorb the $k^0$ and $k^2$ terms generated in $\Delta T^{C_{42}}$ above. (Since we use a cutoff regulator, $\Delta \CLO$ and $\Delta \Ctwo$ must also absorb additional divergences that their insertion itself generates.) 

 The $J_3 \Ctwo \Cft k^4$ piece of  \eqref{eq:pertfoshift1}  can be absorbed into the effect of $\Cfo$ on the on-shell amplitude. 
 Proceeding as we did for $C_{42}$, we compute the  first-order shift due to $C_{41}$, and find:
\begin{align}
\Delta T^{C_{41}}(k,k;E) &= \frac{2 \Cfo g^2(k)}{[(\Ctwo I_2 - 1)^2 - \CLO I_0 - \Ctwo^2 I_0 I_4]^2}
    [(\CLO^2 J_3^2 + \CLO J_5 + \CLO \Ctwo J_3 J_5 \notag \\
    & \quad \null + 2 \Ctwo^2 J_5^2 - \Ctwo^3 J_3 J_5^2 + \Ctwo J_7 - 2\Ctwo^2 J_3 J_7 + \Ctwo^3 J_3^2 J_7)
    \notag \\
    & \quad \null + (3\CLO J_3 - \CLO \Ctwo J_3^2 + 4 \Ctwo J_5 - 2 \Ctwo^2 J_3 J_5) k^2 + (2 - \Ctwo^2 J_3) k^4]  \;. 
     \label{eq:pertfoshift3}
\end{align}

Combining 
(\ref{eq:pertfoshift1}), (\ref{eq:pertfoshift2}), and (\ref{eq:pertfoshift3}) then shows we can satisfy renormalization conditions that these perturbations do not change the zeroth- or second-order pieces of the amplitude (\ref{eq:Tmat-leading-onshell}). The terms proportional to $\Cft \Ctwo$ and $\Delta \Ctwo \Ctwo$ can also
be taken to be part of a redefined $\Cfo$. In other words, the total fourth-order shift (for the stated renormalization conditions) should be written in the form:
\begin{align}
  \Delta T^{C_4}(k,k;E) = \frac{g^2(k)}{[(\Ctwo I_2 - 1)^2 - \CLO I_0 - \Ctwo^2 I_0 I_4]^2} \tilde{C}_{41} k^4  \;,
  \label{eq:totalfourthorderrenormalized}
\end{align}
where $\tilde{C}_{41}$ is a somewhat complicated function of the LECs that appear in (\ref{eq:VSR}). For our purposes the key fact about this function is that $\Cfo$ and $\Cft$ only appear in a linear combination. 
This demonstrates that, at least within the short-range model and in perturbation theory, $\Cft$ is degenerate with $\Cfo$. 

Lastly we note that this argument is much simpler in dimensional regularization with minimal subtraction, where $J_3=J_5=J_7=\ldots=0$.
The argument is also easier in the standard pionless EFT power counting, where $C_2$ is also treated in perturbation theory~\cite{Fleming:1999ee,Rupak:1999aa}. In that case the $k^4$ term does not appear when $\Cft (p^2 - p^{\prime \, 2})$ is evaluated perturbatively; only $k^0$ and $k^2$ terms are present in the numerator.

\section*{References}

\bibliographystyle{myunsrt}

\bibliography{thesis_refs,reactionsineft}

\end{document}